\PassOptionsToPackage{unicode}{hyperref}
\PassOptionsToPackage{hyphens}{url}
\PassOptionsToPackage{dvipsnames,svgnames,x11names}{xcolor}
\documentclass[
]{article}
\usepackage{xcolor}
\usepackage[top=10pc,bottom=10pc,left=11pc,right=11pc,heightrounded,top=1in,bottom=1in,left=1in,right=1in]{geometry}
\usepackage{amsmath,amssymb}
\usepackage{iftex}
\ifPDFTeX
  \usepackage[T1]{fontenc}
  \usepackage[utf8]{inputenc}
  \usepackage{textcomp} 
\else 
  \usepackage{unicode-math} 
  \defaultfontfeatures{Scale=MatchLowercase}
  \defaultfontfeatures[\rmfamily]{Ligatures=TeX,Scale=1}
\fi
\usepackage{lmodern}
\ifPDFTeX\else
\fi
\IfFileExists{upquote.sty}{\usepackage{upquote}}{}
\IfFileExists{microtype.sty}{
  \usepackage[]{microtype}
  \UseMicrotypeSet[protrusion]{basicmath} 
}{}
\usepackage{setspace}

\usepackage{longtable,booktabs,array}
\usepackage{calc} 
\usepackage{etoolbox}
\makeatletter
\patchcmd\longtable{\par}{\if@noskipsec\mbox{}\fi\par}{}{}
\makeatother
\IfFileExists{footnotehyper.sty}{\usepackage{footnotehyper}}{\usepackage{footnote}}
\makesavenoteenv{longtable}
\usepackage{graphicx}
\makeatletter
\newsavebox\pandoc@box
\newcommand*\pandocbounded[1]{
  \sbox\pandoc@box{#1}%
  \Gscale@div\@tempa{\textheight}{\dimexpr\ht\pandoc@box+\dp\pandoc@box\relax}%
  \Gscale@div\@tempb{\linewidth}{\wd\pandoc@box}%
  \ifdim\@tempb\p@<\@tempa\p@\let\@tempa\@tempb\fi
  \ifdim\@tempa\p@<\p@\scalebox{\@tempa}{\usebox\pandoc@box}%
  \else\usebox{\pandoc@box}%
  \fi%
}
\def\fps@figure{htbp}
\makeatother

\NewDocumentCommand\citeproctext{}{}
\NewDocumentCommand\citeproc{mm}{%
  \begingroup\def\citeproctext{#2}\cite{#1}\endgroup}
\makeatletter
 \let\@cite@ofmt\@firstofone
 \def\@biblabel#1{}
 \def\@cite#1#2{{#1\if@tempswa , #2\fi}}
\makeatother
\newlength{\cslhangindent}
\newlength{\csllabelwidth}
\newenvironment{CSLReferences}[2] 
 {\begin{list}{}{%
  \setlength{\itemindent}{0pt}
  \setlength{\leftmargin}{0pt}
  \setlength{\parsep}{0pt}
  \ifodd #1
   \setlength{\leftmargin}{\cslhangindent}
   \setlength{\itemindent}{-1\cslhangindent}
  \fi
  \setlength{\itemsep}{#2\baselineskip}}}
 {\end{list}}
\usepackage{calc}

\providecommand{\tightlist}{%
  \setlength{\itemsep}{0pt}\setlength{\parskip}{0pt}}

\usepackage[all,defaultlines=2]{nowidow}

\usepackage{enumitem}
\setlist[1]{labelindent=\parindent}
\setlist[itemize]{leftmargin=*}
\setlist[enumerate]{leftmargin=*}

\setlist[description]{style=unboxed}

\usepackage[titles]{tocloft}

\AtBeginEnvironment{longtable}{\setlist[itemize]{nosep, wide=0pt, leftmargin=*, before=\vspace*{-\baselineskip}, after=\vspace*{-\baselineskip}}}

\usepackage{setspace}

\usepackage[overload]{textcase}
\usepackage[runin]{abstract}

\abslabeldelim{\quad}
\newenvironment{keywords}
{\vskip -3em \hspace{\parindent}\small\sffamily{\sffamily\footnotesize\bfseries\MakeUppercase{Keywords}}\quad}
{\vskip 3em}

\usepackage{titling}
\pretitle{\par\vskip 5em \begin{flushleft}\LARGE\sffamily\bfseries}
\posttitle{\par\end{flushleft}\vskip 0.75em}

\renewcommand{\and}{\end{tabular} \hskip 3em \begin{tabular}[t]{@{\hspace{0em}}l@{}}}
\preauthor{\begin{flushleft}
           \lineskip 1.5em 
           \begin{tabular}[t]{@{\hspace{0em}}l@{}}}
\postauthor{\end{tabular}\par\end{flushleft}}

\predate{}
\postdate{}

\newcommand{\published}[1]{%
   \gdef\puB{#1}}
   \newcommand{\puB}{}

\usepackage{titlesec}
\titleformat*{\section}{\Large\sffamily\bfseries\raggedright}
\titleformat*{\subsection}{\large\sffamily\bfseries\raggedright}
\titleformat*{\subsubsection}{\normalsize\sffamily\bfseries\raggedright}
\titleformat*{\paragraph}{\small\sffamily\bfseries\raggedright}

\titlespacing*{\section}{0em}{2em}{0.1em}
\titlespacing*{\subsection}{0em}{1.25em}{0.1em}
\titlespacing*{\subsubsection}{0em}{0.75em}{0em}

\usepackage[bottom, flushmargin]{footmisc}

\usepackage[font={small,sf}, labelfont={small,sf,bf}]{caption}

\usepackage{pdflscape}
\newcommand{\blandscape}{\begin{landscape}}
\newcommand{\elandscape}{\end{landscape}}

\usepackage{bbm}
\let\origmathbb\mathbb
\renewcommand{\mathbb}[1]{\ifnum\pdfstrcmp{#1}{1}=0 \mathbbm{1}\else\origmathbb{#1}\fi}
\usepackage{xcolor}
\definecolor{coloraccent}{HTML}{483d8b}
\usepackage[section]{placeins}
\usepackage{float}
\usepackage{tabularray}
\usepackage[normalem]{ulem}
\usepackage{graphicx}
\usepackage{rotating}
\UseTblrLibrary{siunitx}
\NewTableCommand{\tinytableDefineColor}[3]{\definecolor{#1}{#2}{#3}}

\makeatletter
\@ifpackageloaded{caption}{}{\usepackage{caption}}
\AtBeginDocument{%
\ifdefined\contentsname
  \renewcommand*\contentsname{Table of contents}
\else
  \newcommand\contentsname{Table of contents}
\fi
\ifdefined\listfigurename
  \renewcommand*\listfigurename{List of Figures}
\else
  \newcommand\listfigurename{List of Figures}
\fi
\ifdefined\listtablename
  \renewcommand*\listtablename{List of Tables}
\else
  \newcommand\listtablename{List of Tables}
\fi
\ifdefined\figurename
  \renewcommand*\figurename{Figure}
\else
  \newcommand\figurename{Figure}
\fi
\ifdefined\tablename
  \renewcommand*\tablename{Table}
\else
  \newcommand\tablename{Table}
\fi
}
\@ifpackageloaded{float}{}{\usepackage{float}}
\floatstyle{ruled}
\@ifundefined{c@chapter}{\newfloat{codelisting}{h}{lop}}{\newfloat{codelisting}{h}{lop}[chapter]}
\floatname{codelisting}{Listing}

\makeatother
\makeatletter
\@ifpackageloaded{caption}{}{\usepackage{caption}}
\@ifpackageloaded{subcaption}{}{\usepackage{subcaption}}
\makeatother
\usepackage{bookmark}
\IfFileExists{xurl.sty}{\usepackage{xurl}}{} 
\hypersetup{
  pdftitle={Barriers to Gender Convergence},
  pdfauthor={Kazuharu Yanagimoto},
  pdfkeywords={Gender Wage Gap, Social Norms, Job Inflexibility, Home
Production},
  colorlinks=true,
  linkcolor={coloraccent},
  filecolor={Maroon},
  citecolor={coloraccent},
  urlcolor={coloraccent},
  pdfcreator={LaTeX via pandoc}}

\usepackage{orcidlink}  

\title{Barriers to Gender Convergence\thanks{I am grateful to my advisor
Nezih Guner for his continued guidance and support. I thank Pedro Mira,
Tom Zohar, and seminar participants at CEMFI Macro Reading Group, Kansai
Labor Workshop, 47th Simposio de la Asociación Española de Economía, and
the 2023 annual meeting of the Society of Economics of the Household for
their useful suggestions. I acknowledge financial support from the Maria
de Maeztu Unit of Excellence CEMFI MDM-2016-0684, funded by
MCIN/AEI/10.13039/501100011033 and CEMFI, and from JSPS KAKENHI Grant
Number 25K23111.}}

\usepackage{etoolbox}
\makeatletter
\providecommand{\subtitle}[1]{
  \apptocmd{\@title}{\par {\vskip 0.25em \large #1 \par}}{}{}
}
\makeatother
\subtitle{The Interactive Effects of Job Inflexibility and Social Norms}

\author{
{\large Kazuharu Yanagimoto~\orcidlink{0009-0007-1967-8304}}%
 \\%
Kobe University \\%
{\footnotesize \url{yanagimoto@econ.kobe-u.ac.jp}} \and
}

\date{}

\usepackage{url}
\begin{document}
\published{\textbf{Tuesday, March 17, 2026}}

\maketitle

\begin{abstract}
This paper investigates the barriers to gender convergence using Japan
as a salient environment to explore the interactive effects of labor
market structures and social norms. I develop a quantitative model of
household labor supply where couples jointly decide their occupations
and working hours. The model features a labor market with inflexible
``regular'' jobs with convex pay schedules and flexible ``non-regular''
jobs, interacting with social norms regarding spousal earnings. The
calibrated model successfully reproduces observed gender gaps in
participation, occupation, and working hours, and explains 48\% of the
gender wage gap. The model also accounts for cross-regional differences
in gender gaps solely through variation in social norms. Counterfactual
simulations show that while increasing job flexibility substantially
reduces wage and occupational gaps, the working hours gap persists due
to the unequal burden of domestic work. Closing this remaining gap
requires policies such as affordable household services. Furthermore,
the model suggests that the effects of structural reforms can depend on
the strength of gender norms, with larger reductions in gender gaps in
more conservative environments.
\end{abstract}
\vskip 3em

\begin{keywords}
\def\sep{;\ }
Gender Wage Gap\sep Social Norms\sep Job Inflexibility\sep 
Home Production
\end{keywords}

\setstretch{1.25}
\vskip -3em \hspace{\parindent}{\sffamily\footnotesize\bfseries\MakeUppercase{JEL Codes}}\quad {\sffamily\small J16; J22; J31}\vskip 3em

\newpage

\section{Introduction}\label{sec-intro}

Despite significant progress in gender equality over the past century,
gender gaps in labor market outcomes persist in many developed
economies. Two forces have emerged as leading explanations. First,
certain occupations feature convex wage schedules that
disproportionately reward long, inflexible hours, creating a steep cost
for workers who need flexibility (\citeproc{ref-goldin2014}{Goldin
2014}; \citeproc{ref-erosa2022}{Erosa et al. 2022}). Second, social
norms regarding gender roles assign a disproportionate share of domestic
responsibilities to women, constraining their labor supply
(\citeproc{ref-kleven2019a}{Kleven, Landais, and Søgaard 2019};
\citeproc{ref-kleven2019b}{Kleven, Landais, Posch, et al. 2019};
\citeproc{ref-kleven2025}{Kleven et al. 2025};
\citeproc{ref-bertrand2015}{Bertrand et al. 2015}). While each force has
been extensively studied, how they interact to generate and perpetuate
gender disparities remains less understood.

This paper argues that these two forces are mutually reinforcing and
quantifies their interaction within a structural model of household
labor supply. The key insight is that convex wage schedules make short
hours costly, while social norms make long hours for women socially
penalized. Together, they create a ``missing middle'' in the choice set
for women: women must choose between a high-commitment job that violates
social norms or a low-commitment job that entails a significant wage
penalty. To capture this mechanism, I develop a model where couples
jointly decide their occupations, working hours, and domestic labor
hours. Unlike Erosa et al. (\citeproc{ref-erosa2022}{2022}), who take
the allocation of home hours as exogenous (assuming women inherently
dedicate more time to household tasks), my model endogenously generates
these gender differences through its home production component and a
social norm penalty when a wife's earnings exceed her husband's.

I bring this framework to Japan, which provides an ideal laboratory for
studying these interactive effects, precisely because both forces are
strong and clearly observable. Japan is characterized by particularly
strong gender norms and labor market institutions that impose a steep
cost on job flexibility. Unlike other high-income countries where gender
gaps have narrowed, Japan's gender gap in earnings remains persistently
high. Remarkably, this gap coexists with high female labor force
participation, as a large share of working women are concentrated in
low-paying, part-time positions (\citeproc{ref-teruyama2018}{Teruyama et
al. 2018}). These disparities cannot be attributed to differences in
educational attainment, as the gender gap in tertiary education has
largely closed.

What explains this pattern? In the Japanese context, ``part-time'' work
is largely synonymous with \textbf{non-regular} employment, a category
defined by contract status rather than hours worked. To explain why
Japanese women disproportionately sort into these low-paying positions
despite high participation rates, I document two forces. First, the
institutional distinction between \textbf{regular} and
\textbf{non-regular} employment creates a stark trade-off between
earnings and flexibility. Regular jobs typically offer permanent
contracts and higher wages but demand long working hours and inflexible
schedules. In contrast, non-regular jobs are characterized by fixed-term
contracts and lower wages but provide significantly greater flexibility
in terms of hours and location.

Second, I examine the role of social norms, specifically the male
breadwinner norm. I document a sharp discontinuity in the distribution
of wives' earnings relative to their husbands' at the 50\% threshold,
suggesting a strong aversion to wives outearning their husbands. This
``gender identity'' norm (\citeproc{ref-bertrand2015}{Bertrand et al.
2015}) reinforces the traditional division of labor, where women assume
the primary responsibility for domestic work. International survey
evidence from the World Values Survey further confirms that these norms
are particularly pronounced in Japan compared with other high-income
countries. Although the model is calibrated to Japan, the underlying
mechanism is general and applies wherever jobs differ in their
flexibility-wage trade-off and social norms shape the division of
domestic labor.

In the model, each partner is endowed with a productivity level, and the
household faces a domestic labor requirement. A couple chooses among
regular employment, non-regular employment, and non-employment. Regular
jobs feature a convex wage schedule that rewards long hours, while
non-regular jobs offer linear compensation but are subject to a
productivity penalty. The model generates realistic patterns of sorting,
where husbands typically hold regular jobs while wives often select
non-regular positions or exit the labor market to fulfill domestic
responsibilities. The model's parameters are calibrated to match
observed correlations between husbands' and wives' wages and working
hours, as well as the joint distribution of couples' earnings.

After calibration, I evaluate the model's performance by comparing its
predictions against empirical gender gaps, outcomes that were
deliberately \emph{not targeted} during the calibration process. The
baseline model successfully explains a substantial proportion of
observed gender disparities: nearly all gaps in participation rates,
occupational choices, and labor hours, along with 48.1\% of the wage
gap. Additionally, the model accurately reproduces both the joint
distribution of couples' occupational choices (across regular,
non-regular, and non-working categories) and the joint distribution of
working hours conditional on these occupational selections.

I further validate the model's mechanism by exploiting regional
variation in social norms across Japanese prefectures. Using a survey on
attitudes toward gender roles, I construct a prefecture-level social
norm score and show that it is strongly correlated with observed gender
gaps in participation, occupation, hours, and wages. I then feed this
variation into the model by adjusting only the social norm parameter
\(\delta\) across prefectures, holding all other parameters fixed at
their nationally calibrated values. The model, driven solely by
differences in social norms, can account for a significant portion of
the cross-regional variation in gender gaps. This exercise provides
external validation that the interaction between social norms and job
inflexibility is indeed the mechanism driving the observed disparities.

What role do job inflexibility and social norms play in perpetuating
these gender gaps, and how can these disparities be mitigated? To
address these questions, I conduct two counterfactual simulations.
First, I simulate a scenario with \textbf{flexible regular jobs}, where
the wage schedule is linear rather than convex. I find that this
structural change substantially reduces gender gaps in participation,
occupation, and wages. However, the gender gap in working hours remains,
as social norms continue to impose a disproportionate domestic burden on
women. Second, I examine the impact of \textbf{outsourcing housework}.
Japanese couples rarely utilize external housework services and
consequently devote substantial time to domestic tasks, particularly
when raising young children. One contributing factor to this pattern is
Japan's highly restricted international migration, which limits the
availability of household labor.\footnote{ Several studies examine the
  impact of low-skilled migration on women's labor supply, including
  Furtado (\citeproc{ref-furtado2016}{2016}); Cortés and Tessada
  (\citeproc{ref-cortes2011}{2011}); Cortés and Pan
  (\citeproc{ref-cortes2019}{2019}).} By allowing households to purchase
affordable household services, this policy eliminates significant
proportions of all gender gaps, including the working hours gap.

Finally, I investigate the interaction between structural reforms and
social norms. I find that the effectiveness of these structural changes,
such as increasing job flexibility or subsidizing outsourcing, is
heterogeneous depending on the strength of social norms. Specifically,
these policies are more effective in reducing gender gaps when social
norms are stronger. This suggests that the gains from structural reforms
may be larger in societies with more conservative gender norms, such as
Japan, where the distortions these reforms address tend to be more
pronounced.

\textbf{Related Literature} This paper contributes to the growing body
of labor and macroeconomic literature examining the relationship between
household responsibilities and gender gaps in the labor
market.\footnote{ For a comprehensive recent review, see Albanesi et al.
  (\citeproc{ref-albanesi2023}{2023}).}

First, this paper relates to the literature linking social norms to
gender gaps. Kleven, Landais, and Søgaard
(\citeproc{ref-kleven2019a}{2019}) and Kleven, Landais, Posch, et al.
(\citeproc{ref-kleven2019b}{2019}) document that a large portion of the
remaining gender inequality is due to ``child penalties,'' and that
these penalties are correlated with gender norms. Bertrand et al.
(\citeproc{ref-bertrand2015}{2015}) argue that gender identity norms,
specifically the aversion to wives earning more than their husbands, are
a key driver of gender gaps in the US. I contribute to this literature
by explicitly modeling how these norms interact with labor market
structures. Specifically, I demonstrate that social norms regarding
gender roles compel women to assume a disproportionate share of domestic
work, while job inflexibility further exacerbates these disparities. By
elucidating these interactions, I provide insights into the mechanisms
driving marriage and child penalties, highlighting how social norms and
flexibility in occupations reinforce each other to generate gender
disparities.

Second, it connects to the literature on job flexibility and non-linear
wages. Goldin and Katz (\citeproc{ref-goldin2011}{2011}) and Goldin
(\citeproc{ref-goldin2014}{2014}) show that certain occupations feature
convex (non-linear) wage schedules that disproportionately reward long,
inflexible hours, and argue that this variation in the returns to
flexibility across occupations is a key driver of occupational gender
segregation and the gender wage gap. This distinction aligns with
empirical evidence showing that part-time workers typically earn less
per hour than their full-time counterparts
(\citeproc{ref-aaronson2004}{Aaronson and French 2004};
\citeproc{ref-ameriks2020}{Ameriks et al. 2020}). Building on these
insights, Erosa et al. (\citeproc{ref-erosa2022}{2022}) models couples'
occupational decisions across jobs with varying degrees of flexibility.
While their framework effectively captures heterogeneous job flexibility
through a streamlined approach, it treats the allocation of home hours
as exogenous, assuming women inherently dedicate more time to household
tasks than men. In contrast, my model endogenously generates these
differences through its home production component. Another relevant
contribution comes from Cubas et al. (\citeproc{ref-cubas2023}{2023}),
who model the concentration of working schedules by incorporating
penalties for absence during peak hours. They argue that women with
children face disproportionate penalties due to their greater household
responsibilities. While their approach models job flexibility using
detailed time-use data, my framework conceptualizes flexibility through
the convexity of wage schedules.

Finally, this paper significantly advances our economic understanding of
gender gaps in Japan. Despite having one of the largest gender
disparities among developed nations, the underlying drivers of Japan's
gender inequality have received limited attention in economic
literature. Onozuka (\citeproc{ref-onozuka2016}{2016}) examines the
partial convergence in Japan's gender wage gap from 1992 to 2002,
arguing that women were systematically displaced from regular to
non-regular employment during this period. Teruyama et al.
(\citeproc{ref-teruyama2018}{2018}) attributes the growth of non-regular
employment in Japan during the 2000s primarily to increased female labor
supply. Additionally, Kitao and Mikoshiba
(\citeproc{ref-kitao2022}{2022}) provides quantitative analysis of how
fiscal policies influence female labor force participation and
occupational choices. While these studies offer partial explanations for
gender disparities in occupation and wages, few have comprehensively
disentangled the structural causes of Japan's gender gaps, and most
notably, they have not adequately addressed the role of social norms. To
my knowledge, this research represents the first comprehensive analysis
that simultaneously explains Japan's gender gaps across four critical
dimensions: labor force participation, occupational selection, working
hours, and wages.

The remainder of this paper is structured as follows: In the next
section, I document key empirical facts about gender gaps in Japan. In
Section~\ref{sec-model}, I develop the baseline model that captures
couples' joint decisions on occupational choices and working hours while
incorporating social norms. Section~\ref{sec-calib} outlines the
calibration methodology. Section~\ref{sec-baseline} presents the
quantitative results of the baseline model and evaluates its performance
in explaining the gender gaps. Section~\ref{sec-varnorm} validates the
model mechanism using regional variation in social norms. In
Section~\ref{sec-counterfactual}, I conduct counterfactual simulations
to explore the mechanisms driving these gaps, specifically focusing on
the roles of job inflexibility, social norms, and the outsourcing of
housework. Section~\ref{sec-concl} summarizes the findings and discusses
their implications.

\section{Stylized Facts}\label{sec-fact}

This chapter presents descriptive evidence on the sources of gender gaps
in the Japanese labor market. I begin by placing Japan's gender gaps in
international perspective, showing that large earnings gaps coexist with
high female labor force participation and that part-time (non-regular)
employment among women is a key driver. Having established these facts,
I introduce the institutional distinction between \textbf{regular} and
\textbf{non-regular} employment and document how the trade-off between
wages and flexibility embedded in this structure shapes women's
occupational choices. Finally, I examine social norms, particularly the
male breadwinner norm, and argue that they are a root cause of the
disproportionate sorting of women into non-regular jobs.

\subsection{Gender Gaps in International
Perspective}\label{gender-gaps-in-international-perspective}

\begin{figure}

\centering{

\includegraphics[width=0.9\linewidth,height=\textheight,keepaspectratio]{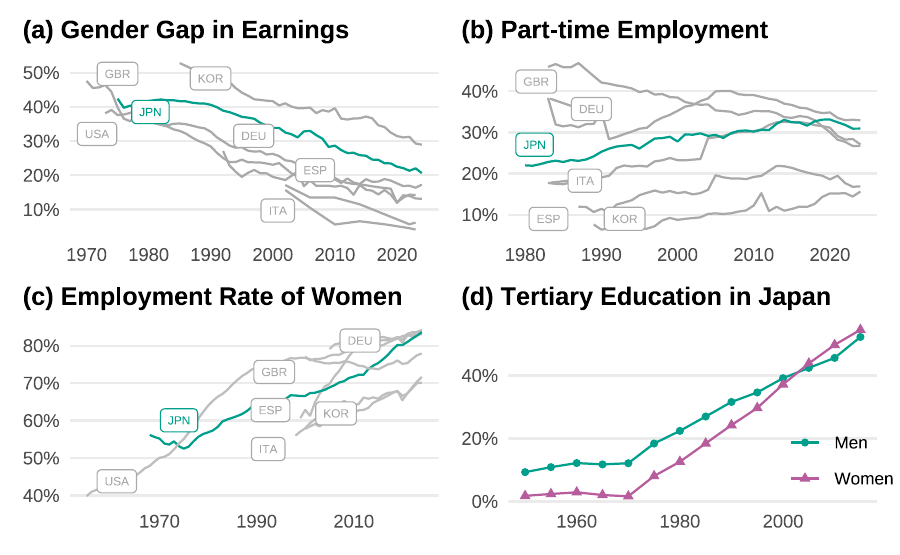}

}

\caption{\label{fig-gaps-comp}\textbf{Labor Market in Japan}. (a) Gender
gap in median earnings of full-time employees relative to male earnings.
(b) Share of part-time employment in total employment. Part-time
employment is defined as usually working less than 30 hours per week in
the main job. (c) Employment rate, calculated as the ratio of employed
persons to the working-age population. (d) Tertiary education attainment
rates in Japan. Sample restricted to ages 25-54. Source from OECD for
(a)-(c) and Barro and Lee (\citeproc{ref-barro2013}{2013}) for (d).}

\end{figure}

Figure~\ref{fig-gaps-comp} summarizes four key facts about Japan's
gender gaps.

\textbf{Japan's gender earnings gap remains large.} Panel (a) compares
the gender gap in median earnings of full-time employees across
high-income countries. While these gaps are gradually closing worldwide,
Japan shows little convergence with international trends, with a gap of
21.3\% in 2022 (the highest among this group after Korea).

\textbf{Female labor force participation is high.} Panel (c) shows that
Japan maintains one of the highest female employment rates among
high-income countries, at 81.8\% in 2022. A large earnings gap thus
cannot be attributed to low female participation.

\textbf{The concentration of women in part-time work is a primary
driver.} Panel (b) reveals that the share of women in part-time
employment has steadily increased in Japan, reaching 31.8\% in 2022,
higher than in other high-income countries. Japan's distinctive
combination of high female participation and high part-time work points
to the sorting of women into low-paying jobs as the central source of
the earnings gap (\citeproc{ref-teruyama2018}{Teruyama et al. 2018}). In
the Japanese context, this ``part-time'' work is largely synonymous with
non-regular employment, a category defined by contract status rather
than just hours.

\textbf{Educational differences do not explain the gap.} Panel (d) shows
that the gender gap in tertiary education has largely closed in Japan.
In 2015, the tertiary attainment rate among adults aged 25-54 is 54.5\%
for women and 52.2\% for men.

These facts raise a central question: why do so many Japanese women end
up in low-paying, part-time (non-regular) jobs despite high
participation and comparable education? The remainder of this chapter
examines two explanations: the institutional structure of the labor
market and social norms regarding gender roles.

\subsection{Data}\label{data}

This paper mainly relies on the Japanese Panel Study of Employment
Dynamics (JPSED). The JPSED is a panel dataset of individual workers
since 2016, with the most recent data wave from 2025.\footnote{ Data
  were distributed one year later. The data in 2024 was distributed in
  2025 and called JPSED2025.} The sample consists of 57,284 individuals
older than 15. This survey contains information on earnings, working
hours, domestic labor hours, and types of jobs. For married individuals,
it also contains information on the spouse's job and earnings.

For the main analysis, I use the pooled data of JPSED2017-2020. The
JPSED2016 does not have information on the domestic labor hours, which
is important for this analysis. The JPSED2021 and later waves may be
affected by the COVID-19 pandemic. The sample includes married men and
women aged 25 to 59 who are in the labor market.

\subsection{Regular and Non-Regular
Workers}\label{regular-and-non-regular-workers}

In the context of Japanese statistics, the terms \textbf{regular} and
\textbf{non-regular} jobs are widely used to categorize employment
types. As this classification is based on each company's internal
categorization, there are no precise legal definitions.\footnote{ See
  Asao (\citeproc{ref-asao2011}{2011}) for a more detailed discussion on
  the definition of regular and non-regular jobs.} However, they are
typically described as follows: A regular worker usually has a permanent
contract, works 40 hours or more at a higher wage, while a non-regular
worker has a temporary contract and works less than 40 hours at a lower
wage.

\begin{figure}

\centering{

\includegraphics[width=0.9\linewidth,height=\textheight,keepaspectratio]{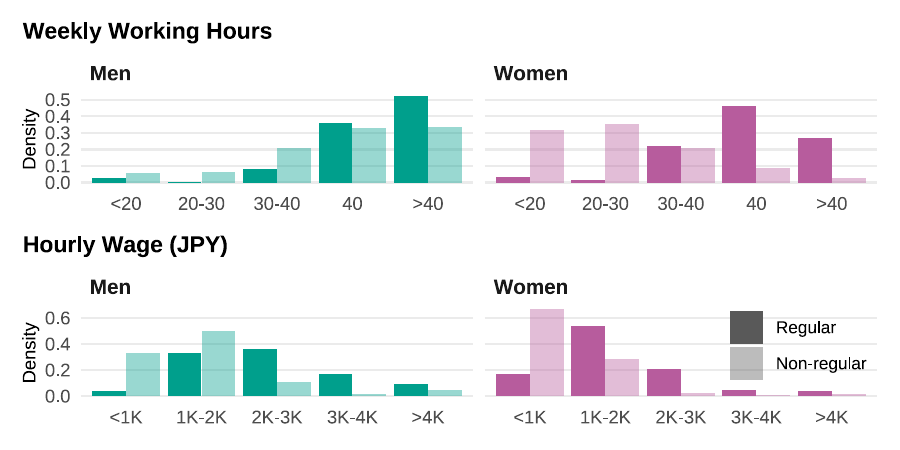}

}

\caption{\label{fig-dist-hours}\textbf{Distribution of Weekly Working
Hours and Hourly Wage}. The data is pooled from JPSED 2017-2020. The
sample includes married men and women aged 25-59.}

\end{figure}

Panel (a) in Figure~\ref{fig-dist-hours} shows that 36.1\% (46.4\%) of
male (female) regular workers work exactly 40 hours per week and 52.3\%
(26.8\%) work more than that. On the other hand, only 33.7\% (3.0\%) of
non-regular male (female) workers work more than 40 hours. In addition,
panel (b) shows that these two occupations differ in hourly wages. While
33.1\% (66.6\%) of non-regular male (female) workers earn a wage of less
than 1000 JPY\footnote{ 1000 JPY \(\simeq\) 5.56 EUR.}, only 4.0\%
(17.0\%) of regular workers do. Note that using the OECD definition of
``less than 30 hours per week in their main job,'' 16.5\% (74.4\%) of
male (female) non-regular workers would be categorized as being in
part-time employment.

\begin{figure}

\centering{

\includegraphics[width=0.9\linewidth,height=\textheight,keepaspectratio]{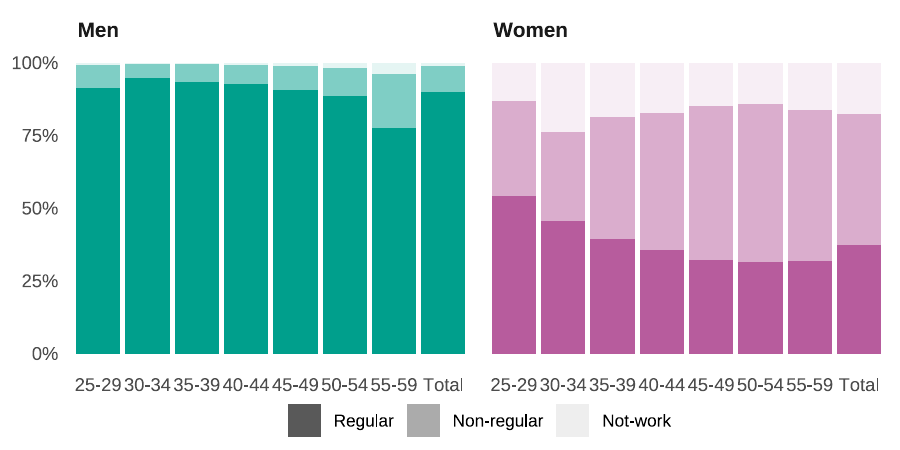}

}

\caption{\label{fig-occ-choice}\textbf{Occupational Choice of Married
Individuals}. The data is pooled from JPSED 2017-2020. The sample
includes married men and women aged 25-59.}

\end{figure}

The gender gap in occupational choices is shown in
Figure~\ref{fig-occ-choice}. We can see clear gender differences in
occupational choices in married individuals. While almost 90\% of men
work as regular workers, less than 30\% of women work as regular
workers, and the share decreases with age. In addition, the proportion
of non-regular workers is much higher among female employees; almost
half of the female employees choose non-regular jobs.

Given the substantial disparities in wages and working hours between
regular and non-regular employment, this divergence in occupational
choice serves as a primary driver of the gender wage gap. Furthermore,
the occupational segregation itself represents a distinct and critical
dimension of gender inequality in the labor market.

\textbf{Differences from Full-time and Part-time Employment} While
economic literature often distinguishes between full-time and part-time
employment, I focus on the distinction between regular and non-regular
jobs. The full-time/part-time dichotomy is typically defined by hours
worked (e.g., the OECD defines full-time as working 30 hours or more per
week). If workers can freely choose their hours, this classification
becomes an endogenous outcome of their labor supply decisions. In
contrast, the regular/non-regular distinction is an exogenous
categorization defined by the employer and employment contract,
representing distinct choice sets available to workers. As discussed in
Section~\ref{sec-job-flex}, these two job types differ fundamentally in
their wage structures and flexibility, presenting workers with a clear
trade-off.

\subsection{Job Flexibility}\label{sec-job-flex}

Why do women choose non-regular jobs? To answer this question, I map
regular and non-regular jobs into non-linear and linear jobs in Goldin
(\citeproc{ref-goldin2014}{2014}), who emphasizes that some jobs have a
highly non-linear (convex) pay structure with respect to working hours,
while others have a nearly linear one. Given a non-linear wage schedule,
a worker can work at a high wage in exchange for long working hours. On
the other hand, a linear job worker can flexibly decide their working
hours since there is no wage penalty for reducing working hours. Hence,
there is a trade-off between job flexibility and wage.

\begin{figure}

\centering{

\includegraphics[width=0.9\linewidth,height=\textheight,keepaspectratio]{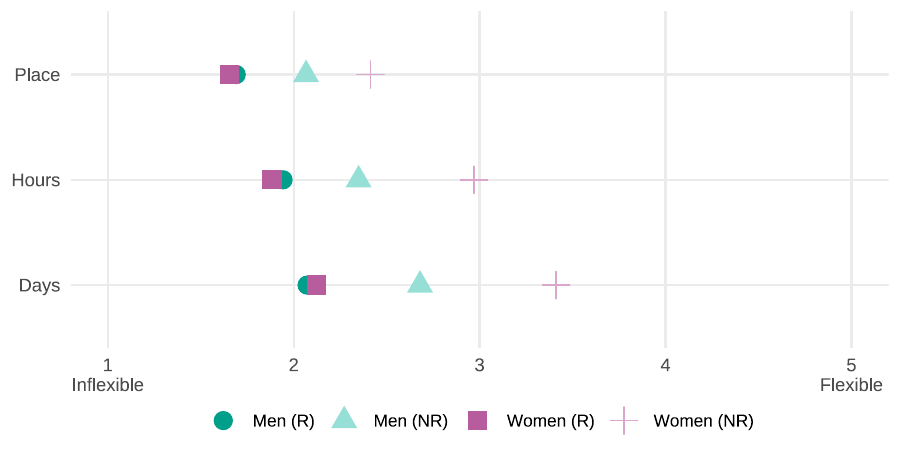}

}

\caption{\label{fig-job-flex}\textbf{Flexibility of Regular and
Non-regular Jobs}. Pooled data from JPSED 2017-2020. The sample includes
married men and women aged 25-59. The figure plots the mean score of job
flexibility (1: Inflexible to 5: Flexible) regarding working days,
hours, and place.}

\end{figure}

Figure~\ref{fig-job-flex} represents a direct measurement of the
flexibility of regular and non-regular jobs, based on survey data. The
questions asked regular and non-regular workers about the flexibility of
their jobs on a 5-point scale: 5 is the highest and 1 is the lowest.
Each point shows the mean job flexibility score in terms of working
days, working hours, and working place. We can see that the regular
worker has less flexibility in all aspects. Also, female non-regular
workers have more flexibility than male regular workers, while female
regular workers face the same inflexibility as male regular workers.

\begin{figure}

\centering{

\includegraphics[width=0.9\linewidth,height=\textheight,keepaspectratio]{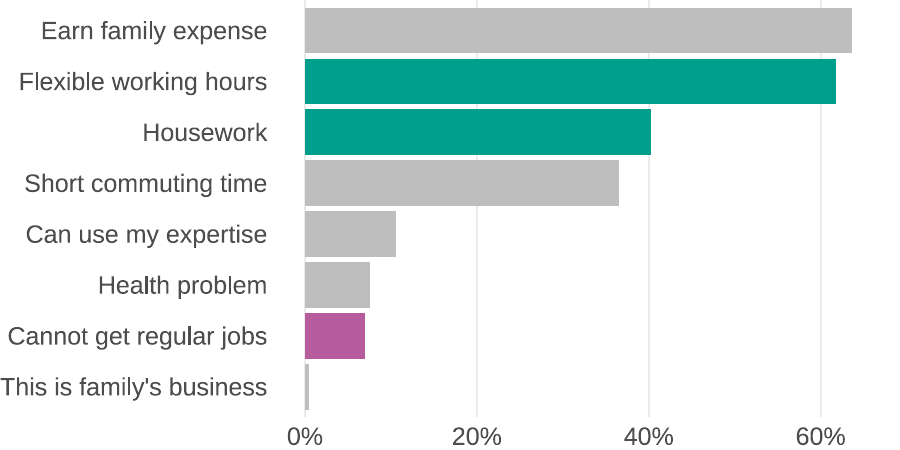}

}

\caption{\label{fig-reasons-nr}\textbf{Reasons Why Women Choose
Non-regular Work}. The data is pooled from JPSED 2017-2020. The sample
includes married women aged 25-59 who have non-regular jobs. Respondents
could select multiple reasons.}

\end{figure}

Finally, if regular jobs require 40 hours of commitment and women have
to allocate a large amount of time to housework, they may choose
non-regular jobs, which are more flexible. Indeed,
Figure~\ref{fig-reasons-nr} supports this argument. This figure shows
the reasons why married women in the non-regular workforce chose their
current job. More than 60\% of the women chose ``job flexibility'' as
the reason and nearly 40\% chose ``housework''. We can also see ``cannot
get regular jobs'' is not the main reason (less than 10\%) why they
chose non-regular jobs.

\begin{figure}

\centering{

\includegraphics[width=0.9\linewidth,height=\textheight,keepaspectratio]{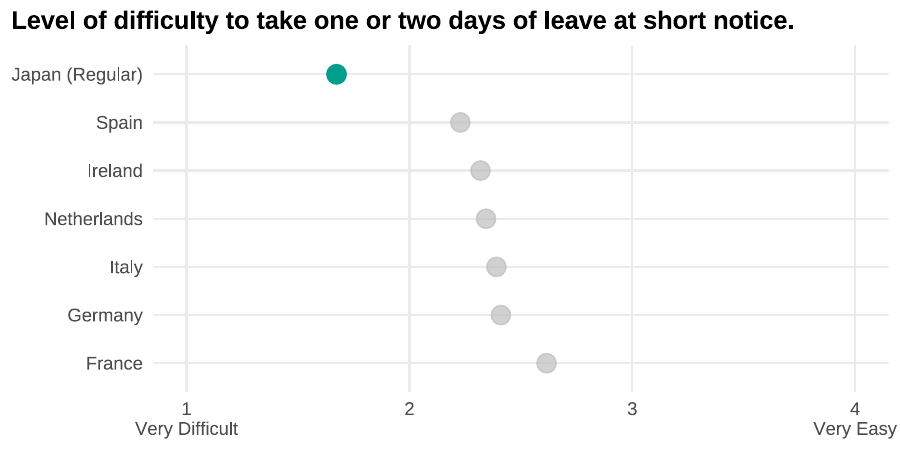}

}

\caption{\label{fig-days-off}\textbf{Flexibility of Working Days}. The
sample includes employed men and women aged 35-49 with children. Data
sources are the Eurostat Labor Force Survey (2019) for European
countries and JPSED for Japan. For Japan, the sample is restricted to
regular workers.}

\end{figure}

While job inflexibility is not unique to Japan, it is more severe than
in other countries. Since there is no survey asking the same question
about job inflexibility in Japan and other countries, as a proxy
measurement, I use a question in JPSED and the Labor Force Survey of
Eurostat. In the JPSED, respondents are asked to rate their agreement
with the statement ``I was able to choose my working days'' on a scale
of 1 to 5. Eurostat includes a question on ``Persons in employment by
level of difficulty to take one or two hours off at short notice'', the
respondents answer on a 1-4 scale. To match the JPSED sample to
Eurostat, I use men and women employed, aged between 35 and 49, and
having children, and rescale the level of difficulty to the 1-4 scale.
Although this is not a direct comparison between Japan and other
countries, Figure~\ref{fig-days-off} indicates that regular workers in
Japan might have more difficulty taking a day off.

\subsection{Social Norms}\label{sec-social-norm}

The trade-off between wages and flexibility documented above applies
symmetrically to men and women: in principle, both can choose between
regular and non-regular jobs. Why, then, do women disproportionately
sort into non-regular employment? In this paper, I argue that social
norms regarding gender roles are a key driving force. Specifically, I
focus on the male \emph{breadwinner} norm: the expectation that husbands
should earn more than their wives. This norm can be interpreted as the
flip side of the expectation that women should bear the primary
responsibility for housework and childcare.

\begin{figure}

\centering{

\includegraphics[width=0.9\linewidth,height=\textheight,keepaspectratio]{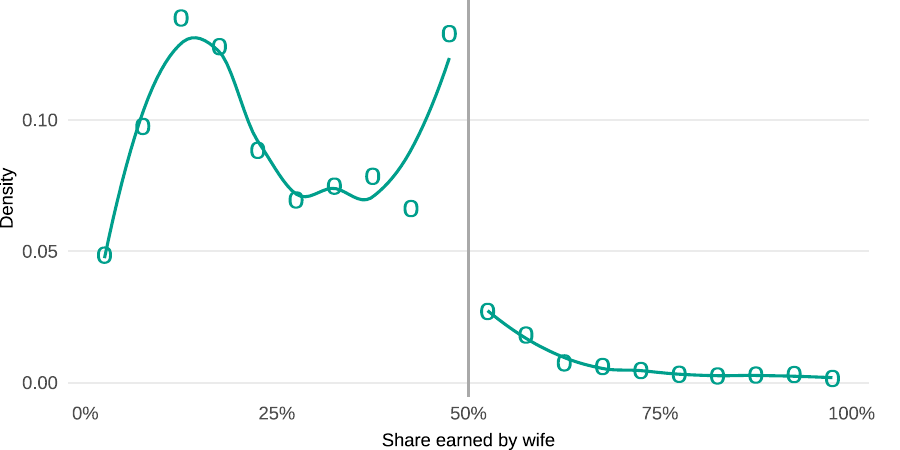}

}

\caption{\label{fig-share-earn}\textbf{Distribution of Relative
Earnings}. The data is pooled from JPSED 2017-2020. The sample includes
married couples aged 25-59. Dual-earner couples only. Each dot
represents the density of couples in a 0.05 relative earnings bin. The
vertical line indicates where the wife's share of earnings is 0.5. The
dashed line is a lowess smoother applied to the distribution, allowing
for a break at 0.5.}

\end{figure}

Bertrand et al. (\citeproc{ref-bertrand2015}{2015}) show that there is a
sharp drop in the distribution of household income share at the point
where wives earn more than their husbands. This type of gender norm
plays a similar role in Japan. Figure~\ref{fig-share-earn} shows the
distribution of the earning share of wives. We observe a clear
discontinuity between the distribution below and above 50\%, which
suggests discontinuous behavior at the point where a wife earns more
than her husband.\footnote{ I conduct robustness check with smaller
  bins. The results, which are similar, are in
  Section~\ref{sec-app-bkp}.} Furthermore, the bunching pattern just
below 50\% supports the strength of this social norm.

\begin{figure}

\centering{

\includegraphics[width=0.9\linewidth,height=\textheight,keepaspectratio]{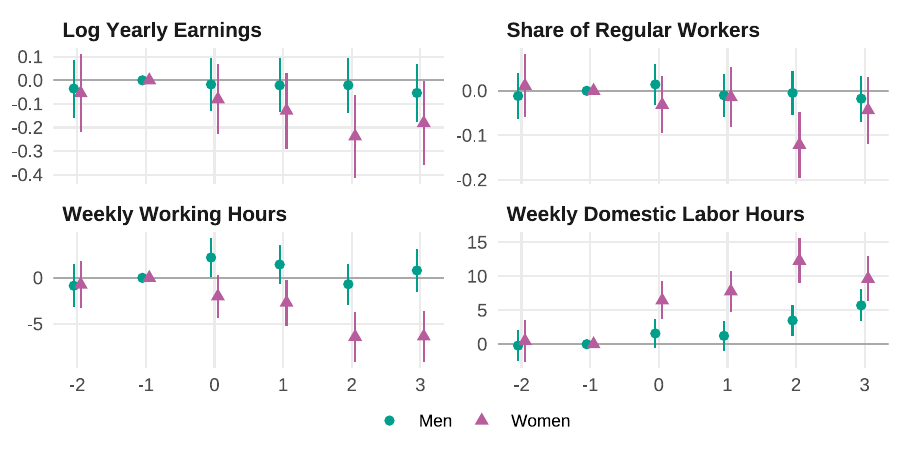}

}

\caption{\label{fig-marriage-penalty}\textbf{Impact of Marriage on Labor
Outcomes}. The data is from JPSED 2016-2025. For sample construction
details, see Section~\ref{sec-app-marriage-penalty}. Each point
represents the estimated coefficient relative to the year prior to
marriage (\(k=-1\)). Error bars indicate 95\% confidence intervals based
on standard errors clustered at the individual level.}

\end{figure}

If there is a penalty for higher wives' earnings, which reflects social
norms, women, upon marriage, will be more likely to choose shorter
working hours, non-regular jobs, or exit from the labor market. In
Figure~\ref{fig-marriage-penalty}, I show the event study of labor
market outcomes since marriage. The sample and the estimation method are
explained in Section~\ref{sec-app-marriage-penalty}. My approach follows
the recent literature that studies the impact of children on gender
gaps, e.g., Kleven, Landais, and Søgaard
(\citeproc{ref-kleven2019a}{2019}), but focuses on marriage rather than
the childbearing event.\footnote{ For a set of European countries,
  {Berniell et al.} (\citeproc{ref-berniell2022}{2022}) estimate the
  separate effects of marriage and childbearing on women's labor market
  outcomes, and find a relatively small role for marriage. Kleven et al.
  (\citeproc{ref-kleven2025}{2025}) compute the marriage and child
  penalty on employment rate for Brazil, China, Japan, Mauritius,
  Rwanda, Sweden, the United Kingdom, and Zambia. Herold and Wallossek
  (\citeproc{ref-herold2023}{2023}) calculates marriage earning gaps
  with German administrative data.}

Since JPSED is an individual survey for workers, I cannot observe
individuals who exit from the labor market after marriage. Thus, I focus
on individuals who are employed before and after marriage. From the
figure, we can see a significant decline in earnings, working hours, and
the share of regular workers only for women, and a significant increase
in domestic labor hours.

If those who exited the labor market were included, the results would
likely be reinforced: labor market exit implies zero earnings, zero
working hours, and no regular employment, along with a likely increase
in domestic labor hours. Therefore, the observed penalties in earnings,
regular employment share, and working hours, as well as the increase in
domestic work, would be even more pronounced.

\begin{figure}

\centering{

\includegraphics[width=0.9\linewidth,height=\textheight,keepaspectratio]{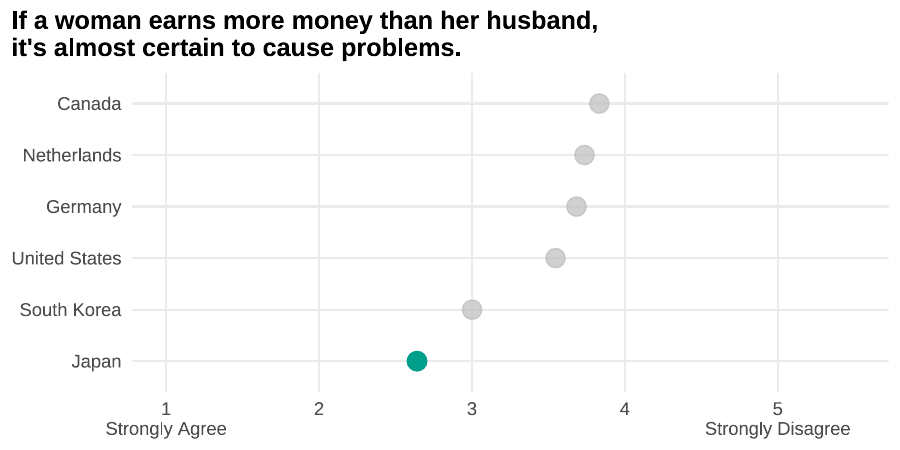}

}

\caption{\label{fig-social-norm}\textbf{Social Norms on Wives'
Earnings}. Each data point represents the mean agreement score by
country (1: Strongly Agree to 5: Strongly Disagree). Source: World Value
Survey Wave 7 (2017-2022).}

\end{figure}

While these social norms are observed in many countries, they are
particularly strong in Japan. The World Values Survey includes a
question asking, ``If a woman earns more money than her husband, it's
almost certain to cause problems.'' Respondents answer on a 1-5 scale,
where 1 indicates ``strongly agree'' and 5 indicates ``strongly
disagree''. Figure~\ref{fig-social-norm} plots the mean of the score by
country. Among these high-income countries, Japan has the strongest
social norms.

\section{Model}\label{sec-model}

\subsection{Overview}\label{overview}

The model economy is populated by married couples that consist of a male
and a female, denoted by \(g \in \{m, f\}\). A couple decides its
occupational choices and the allocation of market and domestic labor
hours to maximize their joint utility. The occupations can be regular,
\(R\), or non-regular, \(NR\). Not working is denoted by \(NW\). If an
individual works, they also decide on their market hours, \(h_m\) and
\(h_f\). Each individual is endowed with one unit of time and a
requirement of joint domestic labor hours \(D\). This requirement
differs across households, and each household decides their domestic
labor hours \(d_m\) and \(d_f\), satisfying \(D = \zeta(d_m, d_f)\).
Each individual is endowed with a productivity (ability) level, denoted
by \(a_{g}\). These ability levels are drawn from a joint distribution
for the couple.

\subsection{Model Components}\label{model-components}

\textbf{Utility Function}

\[
u(c, h + d) = \log c - \phi \frac{(h + d)^{1+\gamma}}{1+\gamma},
\]

where \(c\) is consumption, \(h + d\) is the total labor hours (market
plus domestic), \(\phi > 0\) is a weight on disutility from labor, and
\(\gamma > 0\) captures the curvature of disutility from labor.

\textbf{Productivity} I also assume that the set of productivities
\((a_{m}, a_{f})\) varies across couples. In particular,
\((a_{m}, a_{f})\) is drawn from a log-normal distribution.

\textbf{Convex wage schedules} Following Goldin
(\citeproc{ref-goldin2014}{2014}) and Erosa et al.
(\citeproc{ref-erosa2022}{2022}), regular and non-regular jobs in the
model differ in how hours worked, \(h_m\), map into effective labor
input that determines earnings. For regular jobs, effective labor input
is a convex function of hours worked, i.e., the longer an individual
works, the higher her effective labor input. This creates incentives to
work longer hours since there is an implicit penalty for working short
hours. In contrast, the relation between hours worked and effective
labor input is linear for non-regular jobs. As a result, if one of the
partners cannot supply long hours, they have an incentive to select a
non-regular occupation.

Figure~\ref{fig-dist-hours} shows that 36.1\% (46.4\%) of regular male
(female) workers work exactly 40 hours per week and that 52.3\% (26.8\%)
work more than that. This suggests that 40 hours of work per week is a
standard requirement for regular workers and makes it difficult for
women to work as regular workers.

To capture these two features, i.e., linear vs.~non-linear jobs and a
concentration of 40 hours per week, I assume that the production
function for regular (non-linear) jobs is given by

\[
e(h, a, j = R) = \begin{cases}
a h^{1 + \theta} & \text{ if } h \le \overline{h} \\
a \overline{h}^{\theta} h & \text{ if } h > \overline{h}.
\end{cases}
\]

\(\overline{h}\) is set to 40 hours per week. The parameter
\(\theta > 0\) captures the convexity of the earnings curve for regular
jobs. The slope of the earnings after 40 hours is the same as that at 40
hours, i.e., \(a \overline{h}^{\theta}\).

For non-regular jobs I assume that the production function is linear:

\[
e(h, a, j = NR) = \psi a \overline{h}^{\theta} h.
\]

The parameter \(0 < \psi < 1\) represents a wedge that captures the
difference in productivity between regular and non-regular jobs. If a
regular worker changes to a non-regular job and works the same hours
(but more than 40 hours), her earnings are reduced to a fraction
\(\psi\) of the regular earnings.

\begin{figure}

\begin{minipage}[t]{0.50\linewidth}

\raisebox{-\height}{

\pandocbounded{\includegraphics[keepaspectratio,alt={Regular Job}]{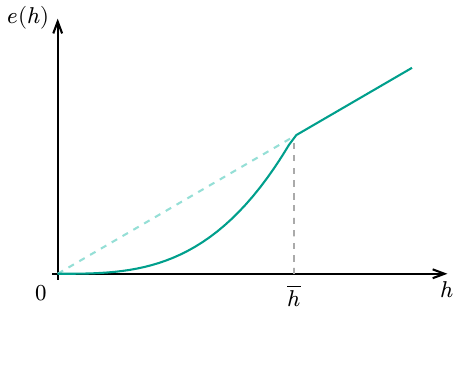}}

}

\subcaption{\label{}Regular Job}
\end{minipage}%
\begin{minipage}[t]{0.50\linewidth}

\raisebox{-\height}{

\pandocbounded{\includegraphics[keepaspectratio,alt={Non-Regular Job}]{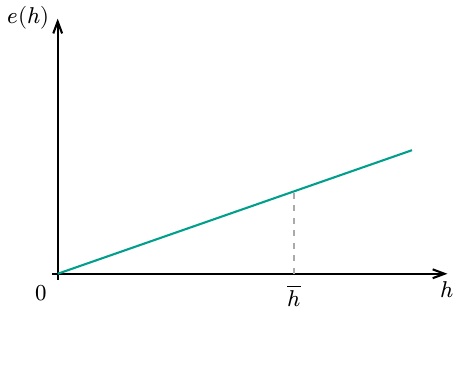}}

}

\subcaption{\label{}Non-Regular Job}
\end{minipage}%

\caption{\label{fig-convex-wage}\textbf{Convex Wage Schedule for Regular
Jobs vs.~Linear Schedule for Non-Regular Jobs}.}

\end{figure}

The wage schedules for regular and non-regular jobs are illustrated in
Figure~\ref{fig-convex-wage}. Since \(\theta > 0\), regular workers have
an incentive to work longer hours. It also means that if one of the
spouses chooses a regular job and works long hours, the other may want
to work short hours and select a non-regular job.

\textbf{Domestic Labor Hours Requirement} I assume that each household
is given a domestic labor hours requirement \(D\) and chooses the
allocation of \(d_m\) and \(d_f\). \(D\) is heterogeneous across
households, which reflects the fact that the amount of domestic labor
hours differs depending on the number of children and their ages. In
particular, I assume that \(D\) is drawn from a cumulative distribution
\(F_{D}\).

To satisfy the requirement, couples need to allocate their domestic
labor hours according to the following joint domestic production
function:

\begin{equation}\protect\phantomsection\label{eq-fn-domestic}{
\zeta(d_m, d_f) := \left(d_m^\xi + d_f^\xi\right)^{\frac{1}{\xi}} = D.
}\end{equation}

If domestic labor hours are perfect substitutes, i.e., \(\xi = 1\), the
couple will allocate all domestic labor hours to the partner with the
lower opportunity cost (wage). However, in reality, couples tend to
share domestic labor hours even when there is a large difference in
wages between the partners. To capture this feature, I assume that
\(\xi < 1\), which implies that domestic labor hours are imperfect
substitutes.

\subsection{Household Problems}\label{household-problems}

Household decisions are made in two steps. First, given productivities
\((a_m, a_f)\) and domestic labor hours requirement \(D\), the couple
chooses their occupational choices \((j_m, j_f)\) to maximize their
expected joint utility. Second, given the occupational choices, they
decide the allocation of market and domestic labor hours to maximize
their joint utility.

\textbf{Household Allocations}

Given their occupational choices \((j_m, j_f)\), the couple chooses
market hours \((h_m, h_f)\) and domestic labor hours \((d_m, d_f)\) to
maximize its joint utility:

\[
U^{j_m, j_f} = \max_{c_m, c_f, h_m, h_f, d_m, d_f} u(c_m, h_m + d_m) + u(c_f, h_f + d_f) - \delta \cdot \mathbb{1}\{e_f > e_m\},
\]

subject to

\[
\begin{aligned}
c_m + c_f & = e(h_m, a_m, j_m) + e(h_f, a_f, j_f), \\
D & = \zeta(d_m, d_f).
\end{aligned}
\]

The last term in the joint utility, which is motivated by
Figure~\ref{fig-share-earn}, represents the utility cost from the wife
earning, \(e_f = e(h_f, a_f, j_f)\), more than the husband,
\(e_m = e(h_m, a_m, j_m)\). \(\delta\) captures the social norm that the
husband should earn more than the wife (breadwinner norm).

It is worth noting that the social norm in the model is formulated as a
penalty on \emph{relative earnings}, not directly on the division of
housework. In practice, these two channels are difficult to disentangle:
a norm that discourages wives from outearning their husbands and a norm
that assigns domestic responsibilities primarily to wives can generate
similar patterns in the data, as both lead women to reduce market hours
and increase domestic hours. The formulation adopted here encompasses
both channels through a single parameter. The key empirical motivation,
the discontinuity in the distribution of relative earnings at the 50\%
threshold (Figure~\ref{fig-share-earn}), is consistent with either
interpretation.

\textbf{Occupational Choices}

The couple chooses their occupational choices \((j_m, j_f)\) to maximize
their expected joint utility:

\[
j_m, j_f = \arg\max_{j_m, j_f} U^{j_m, j_f} + \varepsilon^{j_m, j_f},
\]

where \(\varepsilon^{j_m, j_f}\) represents the idiosyncratic preference
shock for the occupational choice combination \((j_m, j_f)\). I assume
that \(\varepsilon^{j_m, j_f}\) is independently and identically
distributed according to the type-I extreme value distribution. Hence,
the probability that the couple selects occupational choices
\((j_m, j_f)\) is given by the multinomial logit formula:

\[
\text{Pr}(j_m, j_f) = \frac{\exp\left(\frac{U^{j_m, j_f}}{\eta}\right)}{\sum_{j_m' \in \{R, NR, NW\}} \sum_{j_f' \in \{R, NR, NW\}} \exp\left(\frac{U^{j_m', j_f'}}{\eta}\right)},
\]

where \(\eta\) is the scale parameter of the extreme value distribution.

\textbf{Minimum Working Hours}

In the model, I also assume that regular and non-regular jobs have
minimum working hours requirements, denoted by \(\underline{h}_R\) and
\(\underline{h}_{NR}\), respectively. If an individual chooses to work
as a regular (non-regular) worker, they must work at least
\(\underline{h}_R\) (\(\underline{h}_{NR}\)) hours per week. This
captures the fact that firms often set minimum working hours
requirements for their employees. In particular, I set
\(\underline{h}_R\) at 20 hours per week and \(\underline{h}_{NR}\) at
10 hours per week based on the distribution of working hours in
Figure~\ref{fig-dist-hours}.

This condition is also necessary to discipline the model since, without
it, an individual with a job (\(R\) or \(NR\)) working zero hours and an
individual not working (\(NW\)) would yield the same utility, and the
probability of choosing \(R\) or \(NR\) would be ill-defined.

To sum up, the earning schedule is given by

\[
\begin{aligned}
e(h, a, j = R) & = \begin{cases}
0 & \text{ if } h < \underline{h}_{R} \\
a h^{1 + \theta} & \text{ if } h \le \overline{h} \\
a \overline{h}^{\theta} h & \text{ if } h > \overline{h}
\end{cases} \\
e(h, a, j = NR) & = \begin{cases}
0 & \text{ if } h < \underline{h}_{NR} \\
\psi a \overline{h}^{\theta} h & \text{ if } h \ge \underline{h}_{NR}
\end{cases} \\
e(h, a, j = NW) & = 0.
\end{aligned}
\]

\section{Calibration}\label{sec-calib}

\subsection{Calibration Strategy}\label{sub-dist}

I assume that productivity levels \((a_{m}, a_{f})\) are drawn from a
multivariate log-normal distribution. I set the mean of log
productivity, \(\mu = \mathbf{0}\), as a normalization. To reduce the
number of parameters, I assume the variance of the log productivity is
the same between males and females. I did not assume gender differences
in productivity levels a priori, which is consistent with little gender
difference in educational attainment (Figure~\ref{fig-gaps-comp}).
Hence, the gender gaps in earnings and labor market outcomes emerge
endogenously from the model structure.

Based on these assumptions, the productivity levels of a couple are
drawn from:

\[
\log \begin{pmatrix}
\alpha_m \\
\alpha_f
\end{pmatrix} \sim
\mathcal{N} \left(
\begin{pmatrix}
0 \\
0
\end{pmatrix},
\begin{pmatrix}
\sigma^2 & \rho \sigma^2 \\
\cdot & \sigma^2
\end{pmatrix}
\right).
\]

The home requirements also differ across couples according to \[
D \sim Beta(\alpha, \beta).
\]

This heterogeneity represents differences in domestic labor hours based
on the number and age of children. One of the advantages of using the
Beta distribution is that it is defined on a finite interval, \((0,1)\),
which can be satisfied by one person or shared by both.

There are two parameters that I set exogenously. Following Erosa et al.
(\citeproc{ref-erosa2022}{2022}), I set the curvature of labor
disutility, \(\gamma = 3\), so that the Frisch elasticity is fixed at a
value of \(\frac{1}{3}\). Additionally, following Knowles
(\citeproc{ref-knowles2013}{2013}), I set the intra-household elasticity
of substitution for domestic labor, \(\xi = \frac{2}{3}\), which
corresponds to an elasticity of 3.\footnote{ Knowles
  (\citeproc{ref-knowles2013}{2013}) estimates the intra-household
  elasticity of substitution for domestic labor to be around 3 using the
  British Household Panel Survey data. He uses a generalized CES
  production function: \[
    \left((1 - \eta_0) d_m^{1-\eta_1} + \eta_0 d_f^{1-\eta_1}\right)^{\frac{1}{1-\eta_1}},
    \] and estimates \(\eta_0 = 0.475\) and \(\eta_1 = 0.33\).
  \(\eta_0\) is highly close to 0.5, indicating that husbands and wives
  equally share domestic labor in terms of efficiency units, which is
  consistent with the functional form of Equation~\ref{eq-fn-domestic}.}

Given these functional assumptions, there are 9 parameters to be
calibrated:

\[
\Pi = \left(\underbrace{\theta, \psi}_{\text{production}}\,,
\underbrace{\eta}_{\text{shock}}\,,
\underbrace{\phi}_{\text{preference}}\,,
\underbrace{\sigma\text{,}\rho\text{,}}_{\text{productivity}} \,\,
\underbrace{\alpha\text{,}\beta\text{,}}_{\text{domestic labor }}\,\,
\underbrace{\delta}_{\text{social norm}}\right).
\]

The set of parameters \(\Pi\) is estimated to obtain the best possible
fit for the model's predictions. Letting \(D_i\) and \(M_i(\Pi)\) denote
the \(i\)-th data target and the model's solution for this target, the
calibration minimizes the following objective function:
\begin{equation}\protect\phantomsection\label{eq-smm}{
\min_{\Pi} \sum_{i} \left[\frac{D_i - M_i(\Pi)}{D_i}\right]^2.
}\end{equation}

\subsection{Moments}\label{moments}

Table~\ref{tbl-calib-base} summarizes the calibrated parameters and
their corresponding target moments. While the model parameters are
jointly estimated, each is associated with the moment it most directly
impacts.

The production parameter \(\theta\) is determined by the share of
regular workers. The relative utility of non-regular jobs influences the
share of non-regular workers, which is captured by the idiosyncratic
shock parameter \(\eta\). The wage penalty parameter \(\psi\) is pinned
down by the log wage differential between regular and non-regular jobs,
while the labor disutility \(\phi\) is calibrated to the mean working
hours of male regular workers.

The productivity distribution parameters (\(\sigma\), \(\rho\)) are
targeted to the gender wage gap for regular workers, the dispersion of
log wages, and the spousal correlation in earnings. Parameters governing
domestic labor (\(\alpha\), \(\beta\)) are captured by the mean and
standard deviation of domestic labor hours. Finally, the social norm
parameter \(\delta\) is calibrated to match the share of households
where the wife earns more than the husband.

Importantly, female market outcomes are not targeted, except for the
regular wage gap. Thus, the model's predicted gender gaps emerge
endogenously from the structural asymmetries generated by the social
norm \(\delta\).

\begin{table}

\caption{\label{tbl-calib-base}Calibration Results of Baseline Model}

\centering{

\centering
\begin{talltblr}[         
entry=none,label=none,
note{}={Notes: The first and second columns show the estimated values of the model parameters. The third column lists the calibration targets, while the fourth and fifth columns report their moment values in the data and the baseline model, respectively. The parameters are estimated by minimising the distance between the data and model moments defined by @eq-smm.},
]                     
{                     
width={0.85\linewidth},
colspec={X[0.15]X[0.1]X[0.4]X[0.1]X[0.1]},
hline{2}={1-5}{solid, black, 0.05em},
hline{1}={1-5}{solid, black, 0.1em},
hline{11}={1-5}{solid, black, 0.1em},
column{1}={}{halign=c},
column{3}={}{halign=l},
column{2,4-5}={}{halign=r},
}                     
Parameter & Value & Target & Data & Model \\
$\theta$ & 2.62 & Share of regular workers, males & 0.90 & 0.87 \\
$\eta$ & 0.17 & Share of non-regular workers, males & 0.09 & 0.09 \\
$\psi$ & 0.59 & $\overline{\log w_{m, R}} - \overline{\log w_{m, NR}}$ & 0.64 & 0.68 \\
$\phi$ & 12.00 & $\overline{h_{m, R}}$ & 0.40 & 0.40 \\
$\sigma$ & 0.64 & $sd(\log w_{m, R})$ & 0.62 & 0.63 \\
$\rho$ & 0.53 & $\text{Corr}(\log e_{m, R}, \log e_{f, R})$ & 0.21 & 0.21 \\
$\alpha$ & 0.08 & $\overline{d_{f, R}}$ & 0.22 & 0.22 \\
$\beta$ & 0.43 & $sd(d_{f, R})$ & 0.14 & 0.14 \\
$\delta$ & 0.79 & Share of $e_f > e_m$ & 0.07 & 0.07 \\
\end{talltblr}

}

\end{table}%

\subsection{Estimated Results}\label{estimated-results}

The second column in Table~\ref{tbl-calib-base} shows the estimated
parameter values. By comparing the fourth and the fifth columns, which
are the moments in the data and the model, we can see that the model
matches all the targets reasonably well. The convex wage schedule
parameter \(\theta =\) 2.62 is sufficiently larger than zero, indicating
that regular jobs have a non-linear earnings schedule and penalize short
working hours. The wedge of non-regular jobs \(\psi =\) 0.59 implies
that non-regular jobs pay approximately 40.7\% less than regular jobs
for the same hours worked beyond 40 hours per week.\footnote{ One of the
  reasons why non-regular jobs pay less is that non-regular workers are
  less likely to have job training. According to JPSED,42.8\% (37.7\%)
  of male (female) regular workers have additional opportunities for
  training given by their employers (off-the-job training), while 27.1\%
  (21.6\%) of male (female) non-regular workers do.} The gender
correlation in skills \(\rho\) = 0.53 \(>0\) implies positive
assortative mating. The requirement of home hours has the mean 0.16
\(\left(=\frac{\alpha}{\alpha + \beta}\right)\), which corresponds to
18.0 hours per week in the case where only one spouse provides domestic
labor. The rest of the parameters \(\eta =\) 0.17, \(\sigma=\) 0.64,
\(\psi =\) 0.59, \(\delta =\) 0.79 show reasonable values and contribute
to matching the moments.

\section{Baseline Economy}\label{sec-baseline}

\subsection{Occupational Choices and Hours
Worked}\label{occupational-choices-and-hours-worked}

I show how the model economy performs along dimensions that are not
targeted in the calibration. In Table~\ref{tbl-oc-base}, I present the
occupational choice matrix for husbands and wives. The rows are
husbands' jobs and the columns are wives' jobs, and each cell represents
the share of couples with that occupational combination. Overall, the
model explains the distribution of occupations well. However, the share
of non-working husbands is higher in the model than in the data. This
suggests that the inflexibility of regular jobs (convexity of their
earnings) is excessive or the social norms on gender roles are
insufficient.

\begin{table}

\caption{\label{tbl-oc-base}Occupational Choice in Baseline Model}

\centering{

\centering
\begin{talltblr}[         
entry=none,label=none,
note{}={Notes: The table shows the occupational choice of husbands and wives. The rows represent husbands' occupations, and the columns represent wives' occupations. Each cell reports the share of couples in the corresponding occupational combination.},
]                     
{                     
width={0.9\linewidth},
colspec={X[]X[]X[]X[]},
hline{2}={3-4}{solid, black, 0.05em},
hline{3}={1-4}{solid, black, 0.05em},
hline{2}={2}{solid, black, 0.05em, l=-0.5},
hline{1}={1-4}{solid, black, 0.1em},
hline{11}={1-4}{solid, black, 0.1em},
cell{1-2,4-6,8-10}{3}={}{halign=c},
cell{1-2,4-6,8-10}{4}={}{halign=c},
cell{1}{1}={}{halign=c},
cell{1}{2}={c=3}{halign=c},
cell{2,4-6,8-10}{1}={}{halign=l},
cell{2,4-6,8-10}{2}={}{halign=c},
cell{3,7}{1}={c=4}{font=\bfseries, halign=l},
cell{3,7}{2}={}{font=\bfseries, halign=c},
cell{3,7}{3}={}{font=\bfseries, halign=c},
cell{3,7}{4}={}{font=\bfseries, halign=c},
}                     
& Wife &  &  \\
Husband & Regular & Non-regular & Not-work \\
Data & Data & Data & Data \\
Regular & 0.35 & 0.38 & 0.17 \\
Non-regular & 0.01 & 0.07 & 0.01 \\
Not-work & 0.01 & 0.01 & 0.00 \\
Model & Model & Model & Model \\
Regular & 0.17 & 0.33 & 0.37 \\
Non-regular & 0.04 & 0.04 & 0.02 \\
Not-work & 0.04 & 0.00 & 0.00 \\
\end{talltblr}

}

\end{table}%

Finally, I compare the time allocation of couples in each occupation in
Table~\ref{tbl-hour-base}. The model successfully replicates the
characteristic patterns: First, husbands work longer than wives. Second,
regular workers work longer than non-regular workers.

\begin{table}

\caption{\label{tbl-hour-base}Allocation of Weekly Working Hours of
Baseline Model}

\centering{

\centering
\begin{talltblr}[         
entry=none,label=none,
note{}={Notes: The table shows the allocation of weekly working hours by couple's occupations. Columns 1-2 show the husband's and wife's occupations. Columns 3-4 report the data, and Columns 5-6 report the model predictions.},
]                     
{                     
width={0.9\linewidth},
colspec={X[0.2]X[0.2]X[0.125]X[0.125]X[0.125]X[0.125]},
hline{2}={6}{solid, black, 0.05em},
hline{3}={1-6}{solid, black, 0.05em},
hline{2}={3,5}{solid, black, 0.05em, l=-0.5},
hline{2}={4}{solid, black, 0.05em, r=-0.5},
hline{1}={1-6}{solid, black, 0.1em},
hline{7}={1-6}{solid, black, 0.1em},
cell{1}{1}={}{halign=c},
cell{1}{2}={}{halign=c},
cell{1}{3}={c=2}{halign=c, wd=0.25\linewidth},
cell{1}{4}={}{halign=c},
cell{1}{5}={c=2}{halign=c, wd=0.25\linewidth},
cell{1}{6}={}{halign=c},
}                     
&  & Data &  & Model &  \\
Husband & Wife & Husband & Wife & Husband & Wife \\
Regular & Regular & 44.4 & 39.7 & 45.8 & 34.6 \\
Regular & Non-regular & 45.4 & 23.5 & 42.3 & 14.1 \\
Non-regular & Regular & 37.0 & 39.7 & 23.1 & 39.9 \\
Non-regular & Non-regular & 39.8 & 25.5 & 36.5 & 19.1 \\
\end{talltblr}

}

\end{table}%

\subsection{Gender Gaps}\label{sec-gap-base}

How do gender gaps in the model economy, which are not directly
targeted, compare with the data? Table~\ref{tbl-gap-base} shows the
gender gaps in the aggregate economy. The first column shows the
statistics from the data and the second column shows the simulation
results. The third column shows the ratio of the model to the data
column. The first row, the gender gap in participation, represents the
difference in participation rates across genders. The second row, the
gender gap in occupation, represents the difference in the share of
regular workers. For example, the value 0.53 in the data column
represents the difference in the share of regular workers (90.0 \% for
males and 37.4\% for females). The third and fourth rows are the gender
gaps in log working hours and wages. One of the most interesting
findings is that the model explains almost the entire gender gap in
participation rates (208.4\%), occupational choices (117.7\%), and
working hours (158.5\%). This is consistent with the social norm against
wives' higher earnings. Women have an incentive to reduce their earnings
(to zero) by quitting their jobs, changing their occupations, or
reducing their working hours. In addition, the model explains a
significant proportion of the gender gap in wages (48.1\%).

\begin{table}

\caption{\label{tbl-gap-base}Gender Gaps in Baseline Model}

\centering{

\centering
\begin{talltblr}[         
entry=none,label=none,
note{}={Notes: This table shows gender gap measurements and their values in the data and the model. "Participation" refers to the difference in participation rates between males and females. "Occupation" is the gap in the share of regular workers. "Labor Hours" and "Wage" represent the differences in mean log working hours and mean log hourly wages, respectively. The "Ratio" column reports the ratio of the model gap to the data gap.},
]                     
{                     
width={0.9\linewidth},
colspec={X[]X[]X[]X[]},
hline{2}={1-4}{solid, black, 0.05em},
hline{1}={1-4}{solid, black, 0.1em},
hline{6}={1-4}{solid, black, 0.1em},
column{2-4}={}{halign=c},
column{1}={}{halign=l},
}                     
& Data & Model & Ratio \\
Participation & 0.16 & 0.34 & 208\% \\
Occupation & 0.53 & 0.62 & 118\% \\
Labor Hours & 0.49 & 0.77 & 158\% \\
Wage & 0.76 & 0.37 & 48\% \\
\end{talltblr}

}

\end{table}%

\section{Regional Variation in Social Norms}\label{sec-varnorm}

In the previous section, I demonstrated that the baseline model,
calibrated to the national average, successfully reproduces the key
patterns of gender gaps in the Japanese labor market. To further
validate the model's mechanism, specifically the role of social norms, I
now exploit the regional variation across Japanese prefectures. If the
interaction between social norms and labor market structures is indeed
the primary driver of gender gaps, then regions with stronger
traditional norms should exhibit larger gender disparities, and the
model should be able to replicate these cross-regional differences by
varying only the social norm parameter.

Previous studies, such as Abe (\citeproc{ref-abe2013}{2013}), document
substantial cross-prefecture variation in married women's labor force
participation that cannot be explained by individual characteristics,
suggesting that local factors like social norms play a significant role.
This section explicitly measures these local norms and relates them to
gender gaps.

\subsection{Social Norms Score}\label{sec-varnorm-score}

To construct a measure of social norms at the prefecture level, I use
the ``Survey on Awareness of Women's Participation and Advancement in
Regional Areas''\footnote{This is an author translation. The Japanese
  title is ``Chiiki ni okeru Josei no Katsuyaku ni kansuru Ishiki
  Chōsa''.} conducted by the Gender Equality Bureau, Cabinet Office,
Government of Japan in February 2015. The survey includes a battery of
questions on attitudes toward traditional gender roles in the household
and the workplace.

I choose the following five items from the survey as the basis for
constructing the social norms score:

\begin{enumerate}
\def\labelenumi{\arabic{enumi}.}
\tightlist
\item
  \emph{My ideal family is one in which ``the husband works outside the
  home and the wife takes care of the home.''}
\item
  \emph{In general, I think that ``the husband should work outside the
  home and the wife should take care of the home.''}
\item
  \emph{Housework and child-rearing should be done by women.}
\item
  \emph{Caregiving should be done by women.}
\item
  \emph{When children are small, it is better for the mother not to work
  outside the home.}
\end{enumerate}

The responses are on a 4-point scale (``Agree'', ``Somewhat agree'',
``Somewhat disagree'', ``Disagree''). For each item, I take the
prefecture-level average of the responses and then take the mean across
the five items. Higher scores indicate more conservative attitudes
toward gender roles. I use this prefecture-level score as the proxy for
local social norms in the subsequent analysis.

\subsection{Social Norms and Gender Gaps by
Prefecture}\label{social-norms-and-gender-gaps-by-prefecture}

\begin{figure}

\centering{

\includegraphics[width=0.9\linewidth,height=\textheight,keepaspectratio]{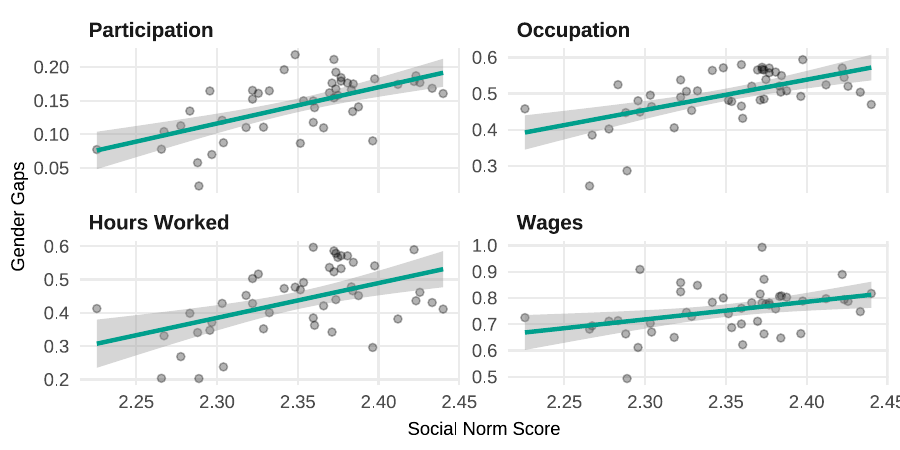}

}

\caption{\label{fig-gap-norm}\textbf{Social Norms and Gender Gaps by
Prefecture}. The figure shows the relationship between social norms
scores (x-axis) and gender gaps in labor market outcomes (y-axis) across
prefectures. Social norms score is defined in
Section~\ref{sec-varnorm-score} and higher score represents more
conservative attitudes toward gender roles. The gender gaps measurement
is defined in Section~\ref{sec-gap-base}. Each point represents a
prefecture, and the solid line is the linear fit (OLS).}

\end{figure}

Figure~\ref{fig-gap-norm} shows the relationship between the social
norms score constructed above and the four gender gaps in labor market
outcomes defined in Section~\ref{sec-gap-base}. Each point represents a
prefecture, and the solid line is the linear fit. We can see a clear
positive relationship in all four panels, indicating that more
conservative social norms are associated with larger gender gaps in
participation, occupation, hours worked, and wages.

\begin{figure}

\centering{

\includegraphics[width=0.9\linewidth,height=\textheight,keepaspectratio]{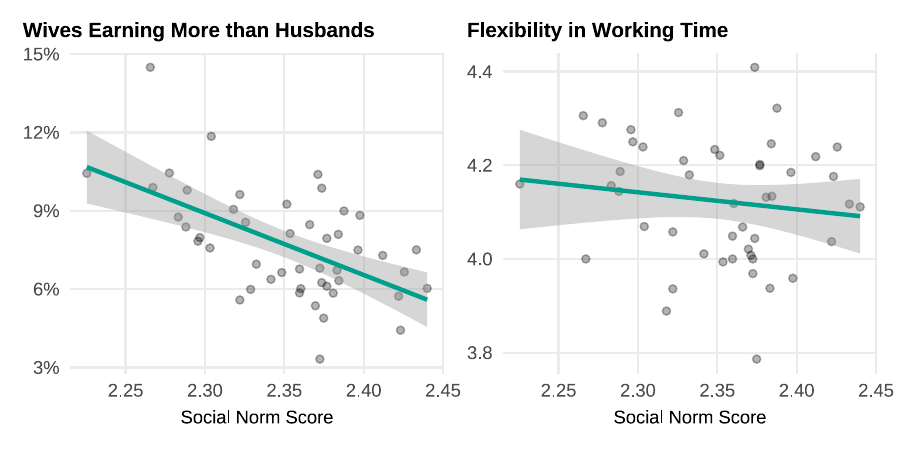}

}

\caption{\label{fig-earnf-flex}\textbf{Social Norms, Share of Wives
Earning More than Husbands, and Work Flexibility}. The left figure shows
the relationship between social norms scores (x-axis) and the share of
wives earning more than husbands (y-axis) across prefectures. The right
figure shows the relationship between social norms scores (x-axis) and
the flexibility in working time (y-axis) among female regular workers
across prefectures. The flexibility measurement is defined in
Section~\ref{sec-job-flex} and higher value represents more flexible
working arrangements.}

\end{figure}

Next, Figure~\ref{fig-earnf-flex} presents two additional relationships
that further illustrate the role of social norms. The left panel shows
the relationship between the social norms score and the share of wives
earning more than their husbands, which is one of the targeted moments
in the baseline economy. The right panel shows the relationship between
the social norms score and the flexibility in working time among female
regular workers, which is a key mechanism in the model as discussed in
Section~\ref{sec-job-flex}. While we can see a clear negative
relationship in the left panel, the right panel shows a weaker
relationship. As discussed in Section~\ref{sec-app-flex-norm}, the
social norm score does not have a significant relationship with most of
the flexibility measures.\footnote{ There is a significant negative
  relationship for the flexibility of female non-regular workers. This
  mostly does not affect the results. This is because the flexibility of
  non-regular jobs is already high in both the model and the data.}
Taken together, these patterns suggest that regional variation in gender
gaps is more directly related to social norms. Motivated by this
observation, I next show that allowing only social norms to vary across
prefectures is sufficient to account for a large share of the observed
regional variation in gender gaps.

\subsection{Model Prediction}\label{model-prediction}

The previous subsection documented that prefectures with more
conservative gender norms exhibit larger gender gaps
(Figure~\ref{fig-gap-norm}). I now ask whether the baseline model can
reproduce this cross-prefecture pattern by varying \emph{only} the
social-norm parameter \(\delta\), holding every other parameter fixed at
its nationally calibrated value \(\widehat{\Pi}_{-\delta}\). In other
words, I construct a model-implied prediction curve on the same axes as
Figure~\ref{fig-gap-norm} (social norm score vs.~gender gaps) and
compare it with the data.

\textbf{Setup.} Let \(\mathbf{g} = (g_1, g_2, g_3, g_4)\) denote the
four gender gap measures (participation, occupation, hours, and wages)
and let \(\mathbf{g}^{\text{data}}_p\) be the vector observed in the
data for prefecture \(p\). I decompose it as

\begin{equation}\protect\phantomsection\label{eq-gap-pref}{
\mathbf{g}^{\text{data}}_p = G(\delta_p; \widehat{\Pi}_{-\delta}) + \mathbf{g}_0 + \Xi_p,
}\end{equation}

where \(G(\cdot)\) maps model parameters to predicted gender gaps and
\(\Xi_p\) is an idiosyncratic error.

\textbf{The role of \(\mathbf{g}_0\).} The term \(\mathbf{g}_0\) is a
\emph{level adjustment} that absorbs the part of the aggregate gender
gap that the model does not explain. It is pinned down from the baseline
economy: letting \(\mathbf{g}^{\text{data}}_{\text{JPN}}\) and
\(\widehat{\delta}\) denote, respectively, the Japan-wide gender gaps
and the nationally calibrated norm parameter, I set

\begin{equation}\protect\phantomsection\label{eq-g0}{
\mathbf{g}_0 := \mathbf{g}^{\text{data}}_{\text{JPN}} - G(\widehat{\delta}; \widehat{\Pi}_{-\delta}).
}\end{equation}

In other words, \(G(\widehat{\delta}; \widehat{\Pi}_{-\delta})\)
corresponds to the ``Model'' column of Table~\ref{tbl-gap-base}, and
\(\mathbf{g}_0\) is the difference between the ``Data'' and ``Model''
columns. Because this wedge is treated as common across prefectures, all
cross-regional variation in the model's predictions is driven entirely
by differences in \(\delta_p\).

\textbf{Linking \(\delta_p\) to the social norm score.} The social norm
score \(s_p\) is a survey-based index, whereas the model is
parameterized by \(\delta\). To place the model's predictions on the
same horizontal axis as the data, I use the share of wives earning more
than their husbands, \(\pi_p\), as a bridge. The model predicts this
share as a function of \(\delta_p\):

\begin{equation}\protect\phantomsection\label{eq-earnf}{
\pi_p = f(\delta_p; \widehat{\Pi}_{-\delta}) + \zeta_p,
}\end{equation}

where \(\zeta_p\) is an idiosyncratic error term. Note that, unlike
Equation~\ref{eq-gap-pref}, no level-adjustment term analogous to
\(\mathbf{g}_0\) appears here. This is because this share is one of the
targeted moments in the calibration, so the baseline model matches the
aggregate data exactly and the residual is zero by construction.

I then relate the score to this share using a linear regression in the
prefecture-level data:

\begin{equation}\protect\phantomsection\label{eq-score-share}{
s_p = \alpha_0 + \alpha_1 \pi_p + \eta_p,
}\end{equation}

where \(\eta_p\) is an error term.

\textbf{Prediction curve.} Combining Equation~\ref{eq-earnf} and
Equation~\ref{eq-score-share} with the gap equation
Equation~\ref{eq-gap-pref} yields a parametric prediction curve indexed
by \(\delta\):

\[
\widehat{s}(\delta) = \hat{\alpha}_0 + \hat{\alpha}_1 f(\delta;\widehat{\Pi}_{-\delta}), \qquad
\widehat{\mathbf{g}}(\delta) = G(\delta;\widehat{\Pi}_{-\delta}) + \mathbf{g}_0.
\]

\begin{figure}

\centering{

\includegraphics[width=0.9\linewidth,height=\textheight,keepaspectratio]{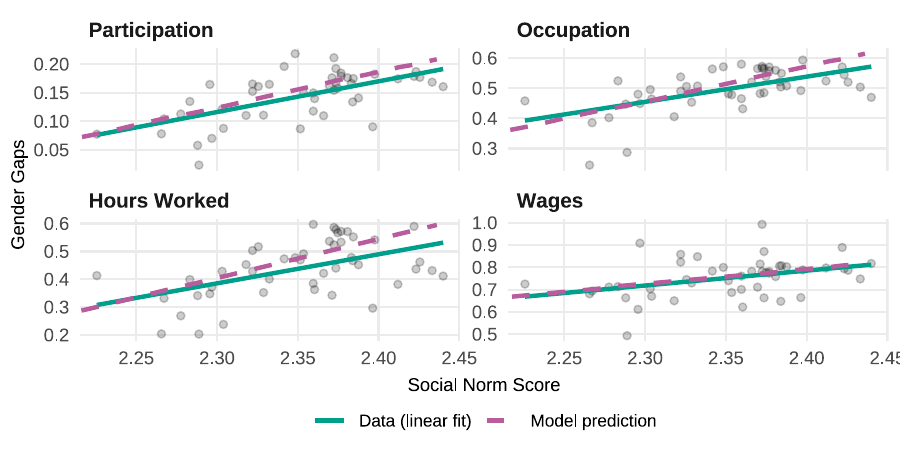}

}

\caption{\label{fig-gap-norm-model}\textbf{Social Norms and Gender Gaps
by Prefecture: Data vs.~Model Prediction}. Points show prefecture-level
gender gaps in the data plotted against the social norm score. The solid
line in each panel is the linear fit in the data (OLS). The dashed line
is the model-implied prediction curve: for each \(\delta\) on the
model's simulated support, the model generates the share of wives
earning more than husbands \(f(\delta;\widehat{\Pi}_{-\delta})\) and the
implied gender gaps \(G(\delta;\widehat{\Pi}_{-\delta})+\mathbf{g}_0\).
To express the model prediction on the score axis, I map
\(f(\delta;\widehat{\Pi}_{-\delta})\) into a predicted score using the
best linear predictor from the prefecture-level regression of the score
on that share.}

\end{figure}

The dashed line in Figure~\ref{fig-gap-norm-model} plots
\((\widehat{s}(\delta), \widehat{\mathbf{g}}(\delta))\) as \(\delta\)
varies over the model's simulated support.

As Figure~\ref{fig-gap-norm-model} shows, the model-implied prediction
curve and the linear fit in the data nearly coincide in each panel. This
close overlap implies that cross-prefecture variation in gender gaps can
be accounted for largely by variation in the model's social-norm
parameter \(\delta\), holding all other calibrated parameters fixed at
their national values. In this sense, regional heterogeneity in social
norms is a sufficient single-index source of variation for organizing
the prefecture-level patterns in participation, occupation, hours
worked, and wages.

\section{Counterfactual Simulations}\label{sec-counterfactual}

\subsection{Counterfactual Scenarios}\label{counterfactual-scenarios}

What accounts for the gender wage gap and part-time work in Japan? In
the model, two factors play a key role in answering these questions:
\textbf{job inflexibility} and \textbf{social norms}. As discussed in
Section~\ref{sec-fact}, Japan has relatively strong social norms on
wives' earnings, and the job inflexibility of Japanese regular workers
is high.

To quantify the impact of these factors, I conduct two counterfactual
simulations.

\textbf{1. Flexible Regular Jobs}: I run simulations with flexible
regular jobs. In the baseline model, regular jobs have a non-linear wage
schedule with a penalty for short hours. In this counterfactual, the
earning schedule \(e(\cdot)\) is set to be linear:

\[
e(a, h, j = R) =\begin{cases}
0 & \text{ if } h < \underline{h}_R \\
a \overline{h}^{\theta} h & \text{ if } h \ge \underline{h}_R
\end{cases}
\]

This removes the pecuniary penalty for working shorter hours in regular
jobs, allowing us to isolate the effect of job inflexibility.

\textbf{2. Outsourcing of Housework}: I consider the outsourcing of
housework. Recent literature emphasizes that the marketization of home
production is a key determinant of female labor supply and the gender
wage gap. For instance, Cortés and Pan (\citeproc{ref-cortes2019}{2019})
show that the availability of market substitutes for household
production allows high-skilled women to work longer hours, thereby
narrowing the gender wage gap. Similarly, Duval-Hernández et al.
(\citeproc{ref-duval-hernandez2023}{2023}) find that cross-country
differences in the marketization of home production explain a
significant portion of the variation in female working hours.
Low-skilled immigration often facilitates this marketization by lowering
the cost of domestic services (\citeproc{ref-furtado2016}{Furtado 2016};
\citeproc{ref-cortes2011}{Cortés and Tessada 2011}).

In contrast, Japan's usage of external household services remains
exceptionally low. According to the Family Income and Expenditure Survey
2021 of the Japanese Statistics Bureau, households of two or more
persons spend only 2.6 euros per \emph{year}.\footnote{I use the
  category ``540 Housekeeping services''.} This is partly due to
restrictive immigration policies.

Suppose households could purchase household services in the market in
Japan. How would this affect gender gaps? To answer this question, I
extend the baseline model as follows.

\[
U^{j_m, j_f} = \max_{h_m, h_f, d_m, d_f} u(c_m, h_m + d_m) + u(c_f, h_f + d_f) - \delta \cdot \mathbb{1}\{e_f > e_m\},
\]

subject to

\[
\begin{aligned}
c_m + c_f + pd &= e_m(h_m, a_{m, j}, j_m) + e_f(h_f, a_{f, j}, j_f), \\
D &= \left(d_m^{\xi} + d_f^{\xi} + d^{\xi}\right)^{\frac{1}{\xi}}.
\end{aligned}
\]

where \(d\) is purchasable housework hours, and \(p\) is its price. This
model allows couples to purchase external household services and to
satisfy the home hours constraint. In other words, it encourages
high-skilled men and women to work more hours in the labor market.

Given the scarcity of housework services, we can consider that the price
of external housework, \(p\), in the benchmark economy is prohibitively
expensive in Japan, so \(d = 0\). Then, I conduct a simulation of the
case where the price of housework services is affordable. In particular,
I set its price to the mean and median wage of a non-regular job in the
benchmark economy (\(p = \psi \overline{h}^{\theta}\)). Here, I assume
that housework services are provided by non-regular workers. The rest of
the parameters are fixed at their calibrated values from the baseline
model.

\subsection{Results}\label{results}

Table~\ref{tbl-occ-cf} compares the choice of occupations between the
baseline, the flexible regular job case, and the outsourcing case. In
the flexible regular job case, we can see an increase in the share of
couples with regular jobs (from 16.5\% to 39.7\%). This is consistent
with the fact that job inflexibility is one of the main reasons for
women choosing non-regular jobs (Figure~\ref{fig-reasons-nr}).

Similar to the case of flexible regular jobs, the outsourcing case also
shows an increase in the share of couples with regular jobs (from 16.5\%
to 52.3\%). This is because outsourcing housework reduces the burden of
domestic labor (Table~\ref{tbl-hours-cf}), thereby allowing wives to
work more in regular jobs.

\begin{table}

\caption{\label{tbl-occ-cf}Comparison in Occupational Choice}

\centering{

\centering
\begin{talltblr}[         
entry=none,label=none,
note{}={Notes: This table shows the fraction of each combination of couples' occupations. The first panel shows the results of the baseline model. The second panel shows the simulation results of linearizing the wage schedule of regular jobs, fixing other parameters. The third panel shows the results of the counterfactual economy with outsourcing of housework.},
]                     
{                     
width={0.9\linewidth},
colspec={X[]X[]X[]X[]},
hline{2}={3-4}{solid, black, 0.05em},
hline{3}={1-4}{solid, black, 0.05em},
hline{2}={2}{solid, black, 0.05em, l=-0.5},
hline{1}={1-4}{solid, black, 0.1em},
hline{15}={1-4}{solid, black, 0.1em},
cell{1-2,4-6,8-10,12-14}{3}={}{halign=c},
cell{1-2,4-6,8-10,12-14}{4}={}{halign=c},
cell{1}{1}={}{halign=c},
cell{1}{2}={c=3}{halign=c},
cell{2,4-6,8-10,12-14}{1}={}{halign=l},
cell{2,4-6,8-10,12-14}{2}={}{halign=c},
cell{3,7,11}{1}={c=4}{font=\bfseries, halign=l},
cell{3,7,11}{2}={}{font=\bfseries, halign=c},
cell{3,7,11}{3}={}{font=\bfseries, halign=c},
cell{3,7,11}{4}={}{font=\bfseries, halign=c},
}                     
& Wife &  &  \\
Husband & Regular & Non-regular & Not-work \\
Baseline & Baseline & Baseline & Baseline \\
Regular & 0.17 & 0.33 & 0.37 \\
Non-regular & 0.04 & 0.04 & 0.02 \\
Not-work & 0.04 & 0.00 & 0.00 \\
Flexible Regular Jobs & Flexible Regular Jobs & Flexible Regular Jobs & Flexible Regular Jobs \\
Regular & 0.40 & 0.23 & 0.25 \\
Non-regular & 0.05 & 0.02 & 0.01 \\
Not-work & 0.03 & 0.00 & 0.00 \\
Outsourcing & Outsourcing & Outsourcing & Outsourcing \\
Regular & 0.52 & 0.23 & 0.15 \\
Non-regular & 0.04 & 0.01 & 0.00 \\
Not-work & 0.03 & 0.00 & 0.00 \\
\end{talltblr}

}

\end{table}%

Table~\ref{tbl-gap-cf} presents the impact of counterfactuals on gender
gaps. The introduction of linear wage schedules substantially reduces
gender disparities, eliminating 32.4\% of the participation gap, 34.7\%
of the occupational gap, and 73.3\% of the wage gap. However, the labor
hours gap remains largely unaffected. Even when the pecuniary penalty
for flexibility is removed, the asymmetry in domestic responsibilities
persists (Table~\ref{tbl-hours-cf}). As long as social norms compel
women to supply the majority of household labor, their opportunity cost
of market work remains high, preventing full convergence in hours.
Consequently, contrary to the prediction in Goldin
(\citeproc{ref-goldin2014}{2014}), the gender gap does not vanish
entirely. The remaining disparities highlight the significant portion of
the gender gap that is attributable to social norms rather than
structural barriers alone.

Outsourcing eliminates significant proportions of all types of gender
gaps (62\% in participation, 49\% in occupation, 45\% in working hours,
73\% in wages). Unlike the case of flexible regular jobs, the gender gap
in labor hours also narrows significantly. This reduction occurs because
outsourcing alleviates the disproportionate burden of domestic work on
women, allowing them to increase their market hours.

\begin{table}

\caption{\label{tbl-gap-cf}Comparison in Gender Gaps}

\centering{

\centering
\begin{talltblr}[         
entry=none,label=none,
note{}={Notes: This table shows gender gaps measurements for baseline, flexible regular job cases, and outsourcing cases. The brackets show the percentage of gaps remaining compared to the baseline.},
]                     
{                     
width={0.95\linewidth},
colspec={X[0.2]X[0.25]X[0.25]X[0.25]},
hline{2}={1-4}{solid, black, 0.05em},
hline{1}={1-4}{solid, black, 0.1em},
hline{6}={1-4}{solid, black, 0.1em},
column{2-4}={}{halign=c},
column{1}={}{halign=l},
}                     
& Baseline & Flexible Regular Jobs & Outsourcing \\
Participation & 0.34 & 0.23 (67.6\%) & 0.13 (38.2\%) \\
Occupation & 0.62 & 0.40 (65.3\%) & 0.32 (51.1\%) \\
Labor Hours & 0.77 & 0.65 (83.6\%) & 0.42 (54.8\%) \\
Wage & 0.37 & 0.10 (26.7\%) & 0.10 (27.3\%) \\
\end{talltblr}

}

\end{table}%

\begin{figure}

\centering{

\includegraphics[width=0.9\linewidth,height=\textheight,keepaspectratio]{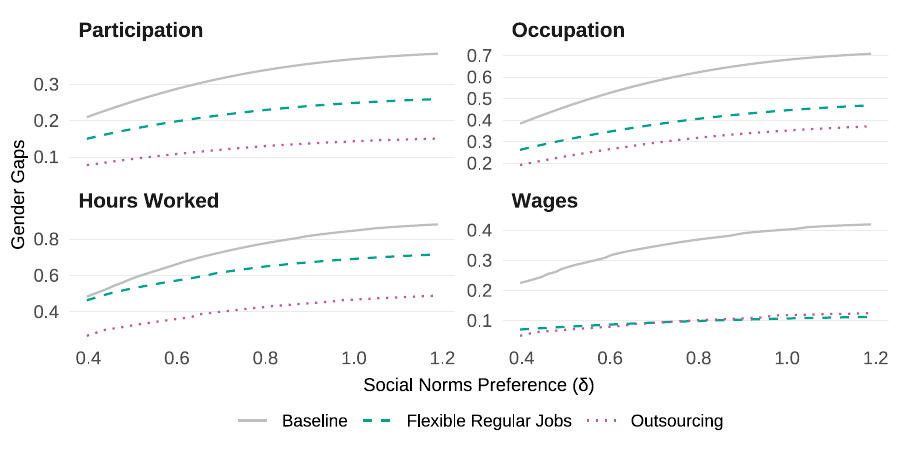}

}

\caption{\label{fig-gap-interaction}\textbf{Interaction between Social
Norms and Counterfactual Scenarios}. The figure shows the relationship
between social norms preference (\(\delta\)) and gender gaps under the
baseline, flexible regular job, and outsourcing scenarios.}

\end{figure}

Finally, I examine how the effect of these counterfactual scenarios
interacts with social norms. Figure~\ref{fig-gap-interaction}
illustrates the heterogeneous effects of the counterfactual scenarios
across different levels of social norm intensity (\(\delta\)). In both
the flexible regular job and outsourcing cases, the reduction in gender
gaps is larger when \(\delta\) is higher. This pattern suggests that the
gains from structural reforms, such as promoting labor market
flexibility or subsidizing household services, may be larger in
environments where gender norms are stronger, such as Japan.

\section{Conclusion}\label{sec-concl}

This paper investigates the interaction between job inflexibility and
social norms in generating gender gaps. I develop a quantitative model
of household labor supply that captures the trade-off between
inflexible, high-paying ``regular'' jobs and flexible, low-paying
``non-regular'' jobs, alongside gender identity norms. The model
successfully reproduces key features of the Japanese labor market,
demonstrating that the combination of inflexible employment structures
and social norms drives women into non-regular employment. Furthermore,
I validate this mechanism by exploiting regional variation across
Japanese prefectures, showing that the model can account for
cross-regional differences in gender gaps solely through differences in
social norms.

Counterfactual simulations show that increasing job flexibility
significantly reduces wage and occupational gaps. However, the gender
gap in working hours persists due to the unequal burden of domestic
work. Policies such as affordable household services are necessary to
close this gap.

Although the analysis focuses on Japan, the mechanism highlighted in
this paper, the interaction between convex wage schedules and gender
norms, is relevant beyond the Japanese context. In any economy where
certain jobs disproportionately reward long hours and social norms
assign domestic responsibilities unequally, the same forces will push
women toward lower-paying, more flexible occupations. The finding that
structural reforms are more effective precisely where norms are
strongest carries a practical implication for policymakers: labor market
reforms in countries with conservative gender norms can yield larger
gains toward gender equality than the same reforms in more egalitarian
settings.

\newpage{}

\appendix
\renewcommand{\thetable}{\Alph{section}.\arabic{table}}
\setcounter{table}{0}
\renewcommand{\thefigure}{\Alph{section}.\arabic{figure}}
\setcounter{figure}{0}

\begin{center}
{\huge\sffamily\bfseries Appendix}
\end{center}
\vspace{1em}

\section{Supplementary Analysis}\label{supplementary-analysis}

\subsection{Specifications of Discontinuity of Relative
Earnings}\label{sec-app-bkp}

\begin{figure}

\centering{

\includegraphics[width=0.9\linewidth,height=\textheight,keepaspectratio]{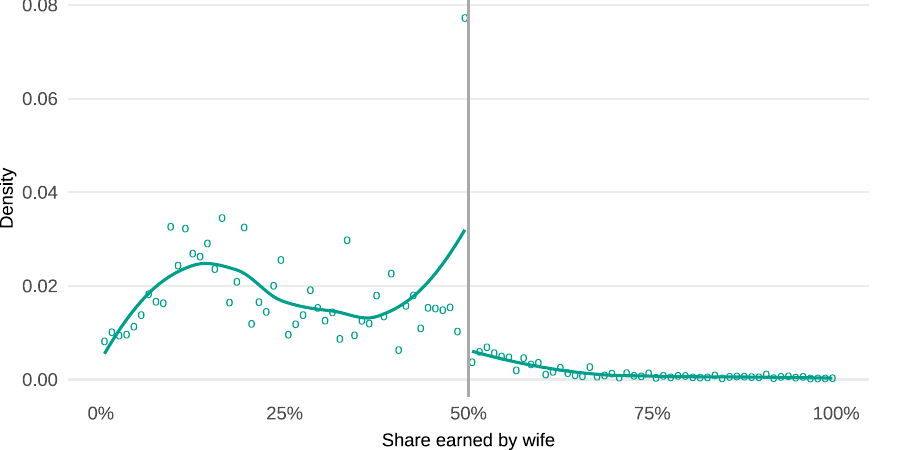}

}

\caption{\label{fig-share-earn-robust}\textbf{Distribution of Relative
Earnings}. The samples are the same as Figure~\ref{fig-share-earn}. The
figure shows the density of the share of earnings between husbands and
wives with 1\% binning.}

\end{figure}

Figure~\ref{fig-share-earn} indicates the gap in the density of share of
earnings between husbands and wives. While this graph simply shows the
discontinuous behavior of couples in terms of earnings, the 5\% binning
may be too coarse. I thus conduct the same analysis with 1\% binning. In
Figure~\ref{fig-share-earn-robust}, we can see discontinuity at 50\% of
the share of earnings in both specifications. We can also see a mass
point at 50\% in the 1\% binning specification, and the density at
exactly 50\% is 0.072. The existence of a mass point at 50\% is also
consistent with the model settings. The utility cost is imposed when a
wife earns strictly more than her husband, and they have an incentive to
earn the same amount. See Kuehnle et al.
(\citeproc{ref-kuehnle2021}{2021}) for a more detailed discussion about
the mass point at 50\% of the share of earnings and the specification of
the discontinuity.

\subsection{Specification of Marriage
Penalty}\label{sec-app-marriage-penalty}

Figure~\ref{fig-marriage-penalty} shows the event study of the following
specification:

\[
y_{it} = \alpha_i + \lambda_t + \sum_{k \neq -1, -\infty} \beta_k \mathbb{1}\left\{e_i + k = t\right\} + \varepsilon_{it},
\]

where \(\alpha_i\) is the individual fixed effect, \(\lambda_t\) is the
year fixed effect, and \(e_i\) is the year of marriage for individual
\(i\). The outcome variable \(y_{it}\) is log of yearly earnings, share
of regular workers, weekly hours worked, and weekly domestic labor
hours.

The sample includes people aged 25 to 59 who are observed in all years
between 2016 and 2023. I use those who got married between 2019 and 2021
as the treatment group to observe the variables \(k = -2, \dots, 3\). As
a comparison group, I use those who never got married during the sample
period.

\subsection{Regional Variation in Social
Norms}\label{sec-app-regional-varnorm}

\begin{figure}

\centering{

\includegraphics[width=0.9\linewidth,height=\textheight,keepaspectratio]{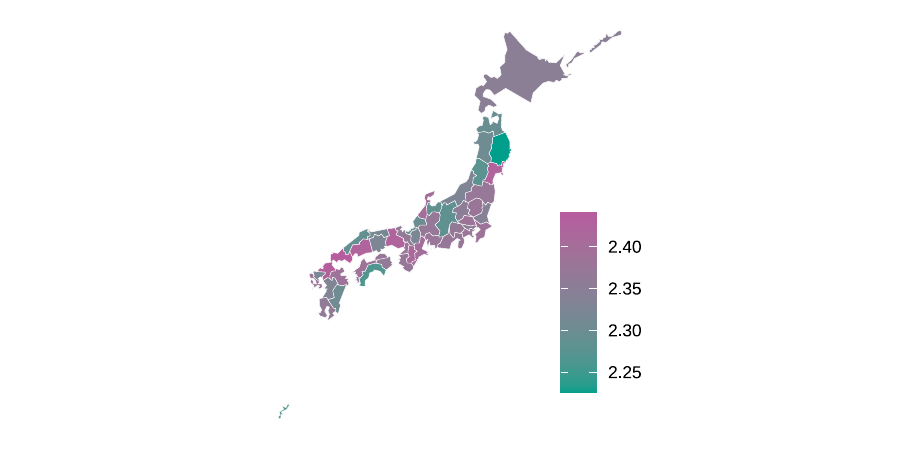}

}

\caption{\label{fig-norm-pref}\textbf{Regional Variation in Social
Norms}. The figure shows the social norm score by prefecture in Japan.}

\end{figure}%

\subsection{Relationship between Flexibility and Social
Norms}\label{sec-app-flex-norm}

\begin{figure}

\centering{

\includegraphics[width=0.9\linewidth,height=\textheight,keepaspectratio]{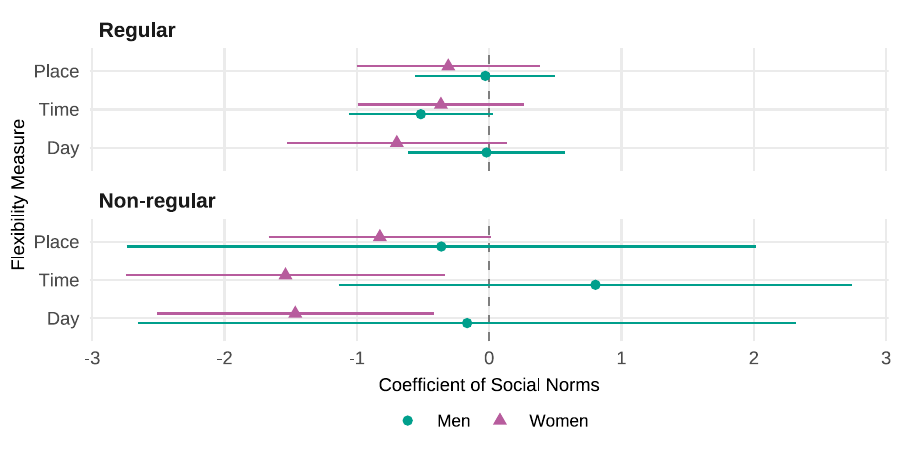}

}

\caption{\label{fig-flex-norm}\textbf{Relationship between Flexibility
and Social Norms}. The figure shows the regression coefficients of the
flexibility measures on the social norms. The error bars indicate the
95\% confidence intervals with robust standard errors.}

\end{figure}

Figure~\ref{fig-flex-norm} shows the regression coefficients of the
flexibility measures on the social norms. The dependent variables are
the flexibility measures (day, time, and place) aggregated by
prefecture, gender, and employment type. The independent variable is the
social norm score in each prefecture. The error bars indicate the 95\%
confidence intervals with robust standard errors.

\subsection{Time Allocation in
Counterfactuals}\label{sec-app-time-alloc-cf}

\begin{table}

\caption{\label{tbl-hours-cf}Working and Home Hours with Outsourcing
\(d\)}

\centering{

\centering
\begin{talltblr}[         
entry=none,label=none,
note{}={Notes: The table compares the allocation of weekly working hours (column 1-2) and domestic labor hours (column 3-4) among the baseline model, the counterfactual economy with flexible regular jobs, and the counterfactual economy with outsourcing.},
]                     
{                     
width={0.9\linewidth},
colspec={X[0.2]X[0.2]X[0.125]X[0.125]X[0.125]X[0.125]},
hline{2}={6}{solid, black, 0.05em},
hline{3}={1-6}{solid, black, 0.05em},
hline{2}={3,5}{solid, black, 0.05em, l=-0.5},
hline{2}={4}{solid, black, 0.05em, r=-0.5},
hline{1}={1-6}{solid, black, 0.1em},
hline{18}={1-6}{solid, black, 0.1em},
cell{1}{1}={}{halign=c},
cell{1}{2}={}{halign=c},
cell{1}{3}={c=2}{halign=c, wd=0.25\linewidth},
cell{1}{4}={}{halign=c},
cell{1}{5}={c=2}{halign=c, wd=0.25\linewidth},
cell{1}{6}={}{halign=c},
cell{3,8,13}{1}={c=6}{font=\bfseries, wd=0.9\linewidth},
cell{3,8,13}{2}={}{font=\bfseries},
cell{3,8,13}{3}={}{font=\bfseries},
cell{3,8,13}{4}={}{font=\bfseries},
cell{3,8,13}{5}={}{font=\bfseries},
cell{3,8,13}{6}={}{font=\bfseries},
}                     
&  & Working Hours &  & Domestic Labor &  \\
Husband & Wife & Husband & Wife & Husband & Wife \\
Baseline & Baseline & Baseline & Baseline & Baseline & Baseline \\
Regular & Regular & 45.8 & 34.6 & 19.0 & 25.6 \\
Regular & Non-regular & 42.3 & 14.1 & 23.7 & 46.4 \\
Non-regular & Regular & 23.1 & 39.9 & 38.4 & 23.9 \\
Non-regular & Non-regular & 36.5 & 19.1 & 27.9 & 41.7 \\
Flexible Regular Jobs & Flexible Regular Jobs & Flexible Regular Jobs & Flexible Regular Jobs & Flexible Regular Jobs & Flexible Regular Jobs \\
Regular & Regular & 37.6 & 24.0 & 26.6 & 36.5 \\
Regular & Non-regular & 40.7 & 16.0 & 24.1 & 43.6 \\
Non-regular & Regular & 31.1 & 29.9 & 32.1 & 32.1 \\
Non-regular & Non-regular & 38.0 & 19.5 & 26.0 & 40.3 \\
Outsourcing & Outsourcing & Outsourcing & Outsourcing & Outsourcing & Outsourcing \\
Regular & Regular & 55.1 & 41.4 & 5.8 & 12.4 \\
Regular & Non-regular & 54.3 & 25.6 & 6.8 & 25.3 \\
Non-regular & Regular & 32.0 & 50.3 & 20.6 & 7.7 \\
Non-regular & Non-regular & 48.8 & 29.7 & 11.9 & 24.0 \\
\end{talltblr}

}

\end{table}%

\newpage{}

\section*{References}\label{references}
\addcontentsline{toc}{section}{References}

\protect\phantomsection\label{refs}


\begin{CSLReferences}{1}{1}
\bibitem[\citeproctext]{ref-aaronson2004}
Aaronson, Daniel, and Eric French. 2004. {``The {Effect} of {Part}-{Time
Work} on {Wages}: {Evidence} from the {Social Security Rules}.''}
\emph{Journal of Labor Economics} 22 (2): 329--252.
\url{https://doi.org/10.1086/381252}.

\bibitem[\citeproctext]{ref-abe2013}
Abe, Yukiko. 2013. {``Regional Variations in Labor Force Behavior of
Women in {Japan}.''} \emph{Japan and the World Economy} 28 (December):
112--24. \url{https://doi.org/10.1016/j.japwor.2013.08.004}.

\bibitem[\citeproctext]{ref-albanesi2023}
Albanesi, Stefania, Claudia Olivetti, and Barbara Petrongolo. 2023.
{``Families, Labor Markets, and Policy.''} In \emph{Handbook of the
{Economics} of the {Family}}, vol. 1. Elsevier.
\url{https://doi.org/10.1016/bs.hefam.2023.01.004}.

\bibitem[\citeproctext]{ref-ameriks2020}
Ameriks, John, Joseph Briggs, Andrew Caplin, Minjoon Lee, Matthew D.
Shapiro, and Christopher Tonetti. 2020. {``Older {Americans Would Work
Longer} If {Jobs Were Flexible}.''} \emph{American Economic Journal:
Macroeconomics} 12 (1): 174--209.
\url{https://doi.org/10.1257/mac.20170403}.

\bibitem[\citeproctext]{ref-asao2011}
Asao, Yutaka. 2011. {``Overview of Non-Regular Employment in {Japan}.''}
\emph{Non-Regular Employment--Issues and Challenges Common to the Major
Developed Countries}, 1--42.

\bibitem[\citeproctext]{ref-barro2013}
Barro, Robert J., and Jong Wha Lee. 2013. {``A New Data Set of
Educational Attainment in the World, 1950--2010.''} \emph{Journal of
Development Economics} 104 (September): 184--98.
\url{https://doi.org/10.1016/j.jdeveco.2012.10.001}.

\bibitem[\citeproctext]{ref-berniell2022}
{Berniell, Inés, Lucila Berniell, Dolores de la Mata, et al.} 2022.
{``Motherhood, Pregnancy or Marriage Effects?''} \emph{Economics
Letters} 214 (May): 110462.
\url{https://doi.org/10.1016/j.econlet.2022.110462}.

\bibitem[\citeproctext]{ref-bertrand2015}
Bertrand, Marianne, Emir Kamenica, and Jessica Pan. 2015. {``Gender
{Identity} and {Relative Income} Within {Households}.''} \emph{The
Quarterly Journal of Economics} 130 (2): 571--614.
\url{https://doi.org/10.1093/qje/qjv001}.

\bibitem[\citeproctext]{ref-cortes2019}
Cortés, Patricia, and Jessica Pan. 2019. {``When {Time Binds}:
{Substitutes} for {Household Production}, {Returns} to {Working Long
Hours}, and the {Skilled Gender Wage Gap}.''} \emph{Journal of Labor
Economics} 37 (2): 351--98. \url{https://doi.org/10.1086/700185}.

\bibitem[\citeproctext]{ref-cortes2011}
Cortés, Patricia, and José Tessada. 2011. {``Low-{Skilled Immigration}
and the {Labor Supply} of {Highly Skilled Women}.''} \emph{American
Economic Journal: Applied Economics} 3 (3): 88--123.
\url{https://www.jstor.org/stable/41288640}.

\bibitem[\citeproctext]{ref-cubas2023}
Cubas, German, Chinhui Juhn, and Pedro Silos. 2023. {``Coordinated {Work
Schedules} and the {Gender Wage Gap}.''} \emph{The Economic Journal} 133
(651): 1036--66. \url{https://doi.org/10.1093/ej/ueac086}.

\bibitem[\citeproctext]{ref-duval-hernandez2023}
Duval-Hernández, Robert, Lei Fang, and L. Rachel Ngai. 2023. {``Taxes,
Subsidies and Gender Gaps in Hours and Wages.''} \emph{Economica} 90
(358): 373--408. \url{https://doi.org/10.1111/ecca.12466}.

\bibitem[\citeproctext]{ref-erosa2022}
Erosa, Andrés, Luisa Fuster, Gueorgui Kambourov, and Richard Rogerson.
2022. {``Hours, {Occupations}, and {Gender Differences} in {Labor Market
Outcomes}.''} \emph{American Economic Journal: Macroeconomics} 14 (3):
543--90. \url{https://doi.org/10.1257/mac.20200318}.

\bibitem[\citeproctext]{ref-furtado2016}
Furtado, Delia. 2016. {``Fertility {Responses} of {High-Skilled Native
Women} to {Immigrant Inflows}.''} \emph{Demography} 53 (1): 27--53.
\url{https://doi.org/10.1007/s13524-015-0444-8}.

\bibitem[\citeproctext]{ref-goldin2014}
Goldin, Claudia. 2014. {``A {Grand Gender Convergence}: {Its Last
Chapter}.''} \emph{American Economic Review} 104 (4): 1091--119.
\url{https://doi.org/10.1257/aer.104.4.1091}.

\bibitem[\citeproctext]{ref-goldin2011}
Goldin, Claudia, and Lawrence F. Katz. 2011. {``The {Cost} of {Workplace
Flexibility} for {High-Powered Professionals}.''} \emph{The ANNALS of
the American Academy of Political and Social Science} 638 (1): 45--67.
\url{https://doi.org/10.1177/0002716211414398}.

\bibitem[\citeproctext]{ref-herold2023}
Herold, Elena, and Luisa Wallossek. 2023. \emph{The {Marriage Earnings
Gap}}.

\bibitem[\citeproctext]{ref-kitao2022}
Kitao, Sagiri, and Minamo Mikoshiba. 2022. {``Why Women Work the Way
They Do in {Japan}: {Roles} of Fiscal Policies.''} \emph{SSRN Electronic
Journal}, ahead of print. \url{https://doi.org/10.2139/ssrn.4054049}.

\bibitem[\citeproctext]{ref-kleven2025}
Kleven, Henrik, Camille Landais, and Gabriel Leite-Mariante. 2025.
{``The {Child Penalty Atlas}.''} \emph{Review of Economic Studies} 92
(5): 3174--207. \url{https://doi.org/10.1093/restud/rdae104}.

\bibitem[\citeproctext]{ref-kleven2019b}
Kleven, Henrik, Camille Landais, Johanna Posch, Andreas Steinhauer, and
Josef Zweimüller. 2019. {``Child {Penalties} Across {Countries}:
{Evidence} and {Explanations}.''} \emph{AEA Papers and Proceedings} 109
(May): 122--26. \url{https://doi.org/10.1257/pandp.20191078}.

\bibitem[\citeproctext]{ref-kleven2019a}
Kleven, Henrik, Camille Landais, and Jakob Egholt Søgaard. 2019.
{``Children and {Gender Inequality}: {Evidence} from {Denmark}.''}
\emph{American Economic Journal: Applied Economics} 11 (4): 181--209.
\url{https://doi.org/10.1257/app.20180010}.

\bibitem[\citeproctext]{ref-knowles2013}
Knowles, John A. 2013. {``Why Are {Married Men Working So Much}? {An
Aggregate Analysis} of {Intra-Household Bargaining} and {Labour
Supply}.''} \emph{The Review of Economic Studies} 80 (3): 1055--85.
\url{https://doi.org/10.1093/restud/rds043}.

\bibitem[\citeproctext]{ref-kuehnle2021}
Kuehnle, Daniel, Michael Oberfichtner, and Kerstin Ostermann. 2021.
{``Revisiting Gender Identity and Relative Income Within Households: {A}
Cautionary Tale on the Potential Pitfalls of Density Estimators.''}
\emph{Journal of Applied Econometrics} 36 (7): 1065--73.
\url{https://doi.org/10.1002/jae.2853}.

\bibitem[\citeproctext]{ref-onozuka2016}
Onozuka, Yuki. 2016. {``The Gender Wage Gap and Sample Selection in
{Japan}.''} \emph{Journal of the Japanese and International Economies}
39 (March): 53--72. \url{https://doi.org/10.1016/j.jjie.2016.01.002}.

\bibitem[\citeproctext]{ref-teruyama2018}
Teruyama, Hiroshi, Yasuo Goto, and Sebastien Lechevalier. 2018.
{``Firm-Level Labor Demand for and Macroeconomic Increases in
Non-Regular Workers in {Japan}.''} \emph{Japan and the World Economy} 48
(December): 90--105. \url{https://doi.org/10.1016/j.japwor.2018.08.006}.

\end{CSLReferences}
\end{document}